\documentclass[a4paper,twocolumn,numberedappendix,twocolappendix,revtex4,iop]{openjournal}
\usepackage[T1]{fontenc}
\usepackage[utf8]{inputenc}
\usepackage{natbib}
\usepackage{amsmath,amssymb}
\usepackage[a4paper,marginpar=10pt,headheight=20pt, right=0.5cm, top=1.5in, left=2cm, bottom=0pt, footskip=0pt, footnotesep=10pt]{geometry}
\usepackage[normalem]{ulem}

\usepackage{booktabs}
\usepackage{graphicx}	% Including figure files
\usepackage{bm}		% Bold maths symbols, including upright Greek
\usepackage{subfigure}
\usepackage[usenames,dvipsnames]{xcolor}
\usepackage[colorlinks,linkcolor=blue,citecolor=blue,urlcolor=blue ]{hyperref}
\usepackage{multirow}
\usepackage{orcidlink}

\usepackage{bbold}
\usepackage{float}

\definecolor{valecol}{rgb}{0,0.5, 1.}

\newcommand\high{\textcolor{red}}

\newcommand\dd{\textrm{d}}

\def\apjl{ApJ Let.}

\def\aap{A\&A}

\def\apjs{ApJS}

\providecommand{\adsurl}[1]{\href{#1}{ADS}}
\providecommand{\eprint}[1]{\href{http://arxiv.org/abs/#1}{#1}}
\usepackage{ifthen}\def\eprinttmp@#1arXiv:#2 [#3]#4@{\ifthenelse{\equal{#3}{x}}{\href{http://arxiv.org/abs/#1}{#1}}{\href{http://arxiv.org/abs/#2}{arXiv:#2} [#3]}}
\renewcommand{\eprint}[1]{\eprinttmp@#1arXiv: [x]@}

\begin{document}

\title{On the full non-Gaussian Surprise statistic and the cosmological concordance between DESI, SDSS and Pantheon+}

\author{
    Pedro Riba Mello,$^{1,2\star}\orcidlink{0009-0005-0325-2400}$
    Miguel Quartin,$^{1,3,4,5,6\sharp}\orcidlink{0000-0001-5853-6164}$
    Bj{\"o}rn Malte Sch{\"a}fer,$^{2,7}\orcidlink{0000-0002-9453-5772}$
    and
    Benedikt Schosser$^{2}\orcidlink{0009-0007-8905-7749}$
    }
\thanks{$^\star$ \href{mailto:peribamello@gmail.com}{peribamello@gmail.com}}
\thanks{$^\sharp$ \href{mailto:mquartin@cbpf.br}{mquartin@cbpf.br}}
\affiliation{
    $^{1}$Instituto de Física, Universidade Federal do Rio de Janeiro, 21941-972, Rio de Janeiro, RJ, Brazil\\
    $^{2}$Zentrum für Astronomie der Universität Heidelberg, Astronomisches Rechen-Institut, Philosophenweg 12, 69120 Heidelberg, Germany \\
    $^{3}$Institut für Theoretische Physik, Universität Heidelberg, Philosophenweg 16, 69120. Heidelberg, Germany \\
    $^{4}$Observatório do Valongo, Universidade Federal do Rio de Janeiro, 20080-090, Rio de Janeiro, RJ, Brazil\\
    $^{5}$PPGCosmo, Universidade Federal do Espírito Santo, 29075-910, Vitória, ES, Brazil \\
    $^{6}$Centro Brasileiro de Pesquisas Físicas, 22290-180, Rio de Janeiro, RJ, Brazil \\
    $^{7}$Interdisziplinäres Zentrum für wissenschaftliches Rechnen, Universität Heidelberg, INF205, 69120 Heidelberg, Germany
    }

\begin{abstract}
    With the increasing precision of recent cosmological surveys and the discovery of important tensions within the $\Lambda$CDM paradigm, it is becoming more and more important to develop tools to quantify accurately the discordance between different probes. One such tool is the Surprise statistic, a measure based on the Kullback-Leibler divergence. The Surprise, however, has been up to now  applied only under its Gaussian approximation, which can fail to properly capture discordance in cases that deviate significantly from Gaussianity. In this paper we developed the \texttt{klsurprise} code which computes the full numerical non-Gaussian Surprise, and analyse the Surprise for BAO + BBN and supernova data. We test different cosmological models, some of which the parameters deviate significantly from Gaussianity. We find that the non-Gaussianities, mainly present in the Supernova dataset, change the Surprise values significantly from its Gaussian approximation, and reveal a borderline $2.0\sigma$ tension in the curved $w$CDM model (o$w$CDM) between the combined Pantheon+ and SH0ES (Pantheon+$\,$\&$\,$SH0ES) data and the dataset which combines SDSS, BOSS and eBOSS BAO. This modest tension is hidden in the Gaussian Surprise approximation. For DESI, the discrepancy with Pantheon+$\,$\&$\,$SH0ES is at the $1.5\sigma$ level for o$w$CDM, but a large $3.4\sigma$ for $\Lambda$CDM. Removing SH0ES data drops  the $\Lambda$CDM significance to $2.6\sigma$.
\end{abstract}

\maketitle
	
%%%%%%%%%%%%%%%%%%%%%%%%%%%%%%%%%%%%
%%%%%%%%%%%%%%%%%%%%%%%%%%%%%%%%%%%%
\section{Introduction}\label{sec:intro}
%%%%%%%%%%%%%%%%%%%%%%%%%%%%%%%%%%%%
\label{ch:Introduction}

In the last 25 years cosmology has developed very successfully, and $\Lambda$CDM became a well established standard model. Over the last few years, however, a comparison of different probes has pointed out some important tensions in the model. In particular, the Hubble constant tension between the values preferred by the Planck satellite data of the Cosmic Microwave Background (CMB)~\citep{Planck:2018vyg} and Supernova observations calibrated locally with Cepheids from the SH0ES program~\citep{Riess:2021jrx}, has been steadily increasing in significance and is now touching the 5$\sigma$ level. This has motivated the community to look into alternative cosmological scenarios on the theory front, and possible systematics on the observational one.

In order to better study tensions, it is essential to quantify and describe them using different model frameworks. In fact, one must quantify the  important question of what it means for two experiments to agree or disagree with each other. Many works have in this sense proposed different concordance and discordance estimators (CDEs). These CDEs should not depend strongly on the functional forms of the analysed distributions. Many of these rely on the Bayesian evidence ratio ${\cal B}$ between datasets~\citep[see, e.g.,][]{Verde:2013wza,Martin:2014lra,Raveri:2015maa}. %
\cite{Hobson:2002zf} discussed how ${\cal B}$ can be used to suggest the need to add hyperparameters to a given model. \cite{Amendola:2012wc} proposed an internal robustness test, which consists of a blind search for systematics by analysing ${\cal B}$ for many random splits of a given dataset \citep[see also][]{Heneka:2013hka,Sagredo:2018ahx}. Some limitations of ${\cal B}$, such as the prior dependence, were addressed in~\cite{Handley:2019wlz, Amendola:2024prl}. See also \cite{Raveri_2019} for a review. The other ingredient needed for quantifying tensions is a scale, with which one compares the chosen metric. The Jeffreys scale is often employed as a way to compare Bayesian evidences, for instance, but it has no fundamental basis and should be seen as empirical. The arbitrariness of Jeffreys' scale can be addressed by assuming a finite model space~\cite{Amendola:2024prl}~\citep[see][for an earlier example]{Castro:2014oja}.

Another commonly used tool to assess agreement is the Kullback-Leibler divergence (KLD)~\cite{Kullback_1951}, also called the relative entropy. It was used in~\cite{March:2011rv} to build a robustness metric which complements the traditional Figure-of-Merit. More recently, it was employed by~\cite{Seehars_2014} where it was used, together with the posterior predictive distribution~\citep[PPD,][]{Huan_2013}, to define the Surprise statistic $S$. It is a quantity which can be interpreted as an information gain from a newer experiment with regard to a previous one~\citep{Lindley1956,Grandis:2015qaa}. The idea is to compare the value $S$ to its distribution, induced from the PPD, and thus compute a $p$-value statistics of the concordance hypothesis between the posteriors.

The KLD is asymmetric, so when comparing different probes which were not in a clear succession, the choice of which is the new and which the previous dataset is not obvious. This asymmetry could be easily solved by symmetrizing both the KLD and the PPD. This would however have the disadvantage of losing the interpretation of information gain between probes, and doubling the (substantial) computational requirements for numerical evaluation. In fact, the computational cost can quickly become the limiting factor, not only for the Surprise but for many of the defined CDEs. Therefore, some simplifying assumptions are often made in the literature. For instance, one can work out analytical solutions for many CDEs assuming Gaussianity. Under the assumption that both likelihood distributions are Gaussian, that the model is a linear function of the parameters and that the priors are also Gaussian or uninformative, the resulting posterior is also Gaussian. This is referred to as the Gaussian Linear Model (GLM). Motivated by the fact that with current CMB  data Gaussianity can be a good approximation within the $\Lambda$CDM model, \cite{Seehars_2014} derived analytical expressions for the Surprise and circumvented the large computational requirements.

Initially, the Gaussian solution for the Surprise statistic was applied to a historical sequence of CMB experiments showing a large Surprise between Wilkinson Microwave Anisotropy Probe~\citep[WMAP,][]{WMAP:2012nax} and the following Planck 2013 data release~\citep{Planck:2013win}. Subsequently, \cite{Seehars_2016} examined the Surprise statistic between WMAP and Planck 2015~\citep{Planck:2015fie}. \cite{Grandis:2016fwl} used the Surprise to quantify discordance between WMAP, Planck, BOSS and Supernova datasets in various cosmological models. A numerical method to evaluate the KLD was presented, but the Surprise statistic distribution was evaluated using the Gaussian solution.

Although larger datasets tend to lead to smaller uncertainties, in general in science we are often testing whether new data is well described by the current paradigm or if new phenomena can be detected, often being described by new degrees of freedom. In the Bayesian context, adding more parameters often make non-Gaussianities larger due to the increase on the confidence intervals, and larger intervals tend to make deviations from linear approximations of the model more evident. This means that Gaussianity in the posteriors can be a constantly eluding goal. In cosmology, in particular, the deviations from Gaussianity in models beyond $\Lambda$CDM are clear for many different observables.

Different methodologies have been proposed to deal with the presence of non-Gaussianities in cosmological posteriors. One approach is to alleviate them by suitable transformation of variables~\citep{Schuhmann:2015dma,Grandis:2016fwl}, although these methods only Gaussianise the posteriors locally, not globally~\citep{Giesel:2020mwj}. Another, is to use the DALI (Derivative Approximation of the Likelihood) method proposed by~\cite{Sellentin:2014zta}, which consists of a higher-order Fisher matrix approach. More recently, a method based on the partition function was also proposed in~\cite{Rover:2022mao}. \cite{Raveri:2021wfz} also proposed a method to quantify discordance based on the Parameter Difference distribution. Their approach does not rely on the Gaussian hypothesis, as they develop methods to evaluate this distribution using either Kernel Density Estimators (KDEs) or Normalizing Flows, a machine learning technique.

For the Surprise statistic, however, these approximations have not been studied, so it remains to be seen whether they are appropriate. In fact, the Surprise is sensitive to the tails of the distributions, a region where the effects of non-Gaussianities is of less importance when considering the posteriors, since in that case the focus is usually on the high-density  regions. Therefore, evaluating the Surprise numerically without assuming approximate Gaussianity may be necessary to properly quantify discordances when using less constraining data and/or when considering models with more degrees-of-freedom than $\Lambda$CDM.

In this paper, we aim to numerically evaluate the Surprise statistic and its distribution and to quantify how much it differs from the case where Gaussianity is assumed. We developed a fully numerical code to compute the Surprise, and apply it to 3 sets of baryonic acoustic oscillation (BAO) and supernova data, discussed below. We show that in most cases results differ substantially from the Gaussian calculations. We will consider models beyond the $\Lambda$CDM, and will devote particular attention to the o$w$CDM model, where both spatial curvature and the dark energy equation of state parameter $w$ are left free to vary. We will also discuss the particular cases $w$CDM (o$\Lambda$CDM), where we held as fixed values $\Omega_{k0}=0$ ($w=-1$), as well as $\Lambda$CDM itself.

\medskip
\subsection{Analyzed Data}

The Baryon Oscillation Spectroscopic Survey~\citep[BOSS,][]{BOSS:2016wmc}
and Extended BOSS~\citep[eBOSS,][]{eBOSS:2019ytm,eBOSS:2020yzd}  were ground-breaking BAO surveys, which in some redshift ranges still contain the highest BAO precision. We will do a combined analysis of BOSS+eBOSS data, complemented at low-redshifts by 6dF~\citep{Beutler:2011hx} and Sloan Digital Sky Survey (SDSS) DR7 galaxies~\citep{Ross:2014qpa}. This combination was previously considered in~\cite{Schoneberg:2019wmt}; here we will refer to it simply as ``BOSS+''.  The Dark Energy Spectroscopic Instrument (DESI) survey~\citep{DESI:2016fyo,Vargas-Magana:2018rbb} is the next-generation step, and will produce a spectroscopic map covering 14000 deg$^{2}$ of the sky, covering the range $z=(0-1.6)$ with a combination of BGS, LRG and ELG galaxies~\citep{DESI:2024uvr}. The 2024 BAO data release (DR1)~\citep{DESI:2024mwx} has significantly extended the redshift coverage of BOSS, and subsequent data releases will far surpass the number of objects from BOSS. In what follows, for simplicity, we will refer to DESI BAO DR1 data simply as ``DESI''.

In terms of supernovae, the Pantheon+ catalogue~\citep{Scolnic:2021amr, Brout:2022vxf} contains 1550 different explosions and is among the most recent and extensive supernova catalogues. We will consider the Pantheon+ both with, and without, the SH0ES~\citep{Riess:2021jrx} calibration. In the latter case, $H_0$ cannot be constrained with supernova.

DESI collaboration results \citep{DESI:2024uvr, DESI:2024mwx} have shown that, in the $\Lambda$CDM model, there is a large tension of more than $3\sigma$ between Pantheon+ and DESI+BBN datasets. In that sense, we will apply both full and Gaussian Surprise statistic to evaluate the pairwise concordance among these latest BAO and supernova datasets, extending the analysis to models beyond $\Lambda$CDM.

%%%%%%%%%%%%%%%%%%%%%%%%%%%%%%%%%%%%
\medskip
\subsection{Choice of priors}
\label{sec:priors}
%%%%%%%%%%%%%%%%%%%%%%%%%%%%%%%%%%%%

For the o$w$CDM model, which is the most general model we consider, we adopt the following broad top-hat priors: $p(h) = {\cal U}[0.3, \,1.0]$, $p(\Omega_{m0}) = {\cal U}[0.05, \,1]$, $p(\Omega_{k0}) = {\cal U}[-1, \,1]$ and $p(w) = {\cal U}[-3, \, -0.4]$. For the other models in which we fix $\Omega_{k0}$ and/or $w$, we leave the priors on the other parameters unchanged. We tested that even broader priors lead to no significant changes in our results. We consider alternative prior choices in Appendix~\ref{app:priors}.

%%%%%%%%%%%%%%%%%%%%%%%%%%%%%%%%%%%%
\bigskip
\section{Comparing distributions and the Surprise statistic}
\label{sec:surprise}
%%%%%%%%%%%%%%%%%%%%%%%%%%%%%%%%%%%%

In this section, we summarize the Surprise statistic and explore the numerical approaches to its calculation.

\subsection{The Gaussian Surprise statistic}

The Kullback-Leibler divergence provides a measurement of the difference between two posterior distributions $p(\Theta|\mathcal{D}_1)$ and $p(\Theta|\mathcal{D}_2)$.\footnote{It is sometimes useful to define the KLD between the posterior and the prior~\citep{Kunz:2006mc}.}
It is a quantity borrowed from Information Theory, its axiomatically defined and reads:
\begin{equation}
    D_{\rm KL}\big[p(\Theta|\mathcal{D}_2)\big\|p(\Theta|\mathcal{D}_1)\big] =\! \int \dd\Theta \, p(\Theta|\mathcal{D}_2) \log \left(\frac{p(\Theta|\mathcal{D}_2)}{p(\Theta|\mathcal{D}_1)}\right)\!.
    \label{eq:kld}
\end{equation}
The relative entropy is given in units of the chosen logarithm base; in this specific case, natural units, or \emph{nats}. A simple division by $\log2$ converts nats into bits. The relative entropy works as a measure of discordance, as it is zero if, and only if, both distributions are equal almost everywhere, i.e.~everywhere except on a subset of measure zero. It is otherwise always positive and the larger its value, the more discordant the distributions are. It is also invariant under reparametrizations.

Since the KLD is an integration over the complete parameter space, it incorporates all information present in both posteriors, and thus accounts for discrepancies in a more complete way than the usual metrics based on comparing marginalized posteriors. The KLD in the Gaussian case has a known analytical solution~\citep{Seehars_2014, Seehars_2016}:
\begin{equation}
    \begin{aligned}
    D_{\rm KL}^{\rm G}\big[p_2\|p_1\big]= \;& \frac{1}{2}\left(\Theta_1-\Theta_2\right)^T \Sigma_1^{-1}\left(\Theta_1-\Theta_2\right) \\
    & \!\!\!\!\! +\frac{1}{2}\left(\operatorname{tr}\left(\Sigma_2 \Sigma_1^{-1}\right)-d-\log \frac{\operatorname{det}\left(\Sigma_2\right)}{\operatorname{det}\left(\Sigma_1\right)}\right) ,
    \end{aligned}
    \label{eq:kld_analytical}
\end{equation}
where $p_1 \equiv p(\Theta|\mathcal{D}_1)$ and $p_2 \equiv p(\Theta|\mathcal{D}_2)$ are the posterior distributions from two different experiments, $\Theta_i$ are the best-fit values, $\Sigma_i$ are the covariance matrices and $d$ is the dimensionality of the parameter space. We also add an upper index G to any quantity for which we use its Gaussian approximation. Using Eq.~\eqref{eq:kld_analytical} and considering the case of a one-dimensional Gaussian distribution, we see that a KLD value of 1 bit corresponds to a shift of about $1.18\sigma$ in central values or a reduction of 68\%  in standard deviation, and a $p$-value of  0.24.

Although assuming Gaussianity can often be a reasonable approximation, there are many known exceptions in cosmology. This is particularly true for extensions of $\Lambda$CDM with a free dark energy equation of state parameter $w$. In particular, many cosmological models are significantly non-linear in the most often used parameters. Moreover, the constant search for new physics frequently entails adding extra weakly constrained parameters to the analysis. These, in turn, can lead to significant non-Gaussianities. In such cases, one can evaluate Eq.~\eqref{eq:kld} numerically performing a Monte Carlo integration:
\begin{equation}
    D_{\rm KL}\big[p_2\|p_1\big] = \frac{1}{N}\sum_{i=0}^N [ \log p(\Theta_i|\mathcal{D}_2) -\log p(\Theta_i|\mathcal{D}_1) ],
    \label{eq:kld_numerical}
\end{equation}
where $\{\Theta_i\}$ are samples from distribution $p(\Theta|\mathcal{D}_2)$. In Appendix~\ref{app:kld_convergence} we provide technical details on the number of samples needed for an accurate computational of the KLD.

In order to define a better motivated scale for the KLD and other relevant quantities, one can resort to the PPD, defined as:
\begin{equation}
    p(\mathcal{D}_2|\mathcal{D}_1) = \int \dd\Theta \,p(\Theta|\mathcal{D}_1)\mathcal{L}(\mathcal{D}_2|\Theta).
    \label{eq:PPD}
\end{equation}
The posterior predictive distribution gives a way to forecast the data distribution of a new experiment $\mathcal{D}_2$ taking as prior information the posterior $p(\Theta|\mathcal{D}_1)$ of a previous one. The PPD also has a closed analytical form when the likelihoods are Gaussian and the data is linear in its parameters. For a more general form of likelihoods and/or model, one can sample the PPD numerically by integrating Eq.~\eqref{eq:PPD} by Monte Carlo integration and using a preferred sampling method:
\begin{equation}
    p(\mathcal{D}_2|\mathcal{D}_1) = \frac{1}{N}\sum_{i=0}^N\mathcal{L}(\mathcal{D}_2|\Theta_i)\,,
    \label{eq:PPD_numerical}
\end{equation}
where $\{\Theta_i\}$ are samples from the posterior $p(\Theta|\mathcal{D}_1)$. As the posterior predictive distribution is a function of data, it has the dimensionality of the dataset. For current type Ia supernovae (SN Ia) catalogues, for instance, one can have a $\sim 2000$ dimensional function, which limits the choice of sampling strategies for the PPD. But considering that most likelihoods are well approximated by Gaussians, the sampling process can be made much simpler and thus feasible.

From the PPD, one can finally derive an expected value for the relative entropy $D_{\rm KL}$ by integrating Eq.~\eqref{eq:kld} over all possible data realizations for $\mathcal{D}_2$:
\begin{equation}
    \langle D_{\rm KL} \rangle = \int \dd\mathcal{D}_2 \, p(\mathcal{D}_2|\mathcal{D}_1)  \, D_{\rm KL}\big[p(\Theta|\mathcal{D}_2)\|p(\Theta|\mathcal{D}_1)\big].
    \label{eq:exp_kld}
\end{equation}
The expected relative entropy also has an analytical solution for the GLM case, which is given by
\begin{equation}
    \langle D_{\rm KL}^{\rm G}\rangle=-\frac{1}{2} \log \frac{{\rm det}\, \Sigma_{2}}{{\rm det}\,\Sigma_{\text {1}}}+\operatorname{tr}\left(\Sigma_{1}^{-1} \Sigma_2 \right) .
    \label{eq:exp_kld_analytical}
\end{equation}
With the definition of both the relative entropy and its expected value, \cite{Seehars_2014} defined the Surprise statistic $S$ as:
\begin{equation}
    S = D_{\rm KL}-\langle D_{\rm KL} \rangle .
    \label{eq:surprise}
\end{equation}

The Surprise $S$ is a simple scalar and its distribution, by definition, has zero mean. The more positive $S$ is, the more different both posteriors are. A negative $S$ means that both posteriors are more compatible than statistically expected. Clearly, in order for the Surprise value to have any meaning, one must compare it to its expected distribution. This distribution is induced by the posterior predictive distribution and is the same as the KLD distribution up to a simple shift factor given by $\langle D_{\rm KL} \rangle$. With $S$ and its distribution, one can compute the statistical significance of the result. In particular, one can compute the $p$-value statistic. As shown in~\cite{Seehars_2014, Seehars_2016}, if both likelihoods are Gaussian and the data is linear in the parameters, the Surprise and its distribution have an analytical solution, which is given by
\begin{equation}
    S^{\rm G}=\frac{1}{2}\left[\Delta \Theta^T \Sigma_1^{-1} \Delta \Theta-\mathrm{tr}(\mathbb{1} \pm \Sigma_2 \Sigma_1^{-1})\right].
    \label{eq:surprise_analytical}
\end{equation}
Inside the trace term, the minus sign is used when $\mathcal{D}_2$ was used to update the posterior $p(\Theta|\mathcal{D}_1)$, while the $+$ sign is used when $p(\Theta|\mathcal{D}_2)$ was derived independently of $\mathcal{D}_1$. The distribution of $S$ follows a generalized $\chi^2$ distribution and can be rewritten as a weighted sum of $\chi^2$ variables $Z_i$ with one degree of freedom, given by
\begin{equation}
    \begin{aligned}
    S^{\rm G} & =\frac{1}{2}\left(\sum_{i=1}^d \lambda_i Z_i-\operatorname{tr}\left(\mathbb{1} + \Sigma_2 \Sigma_1^{-1}\right)\right) \\
    & =\frac{1}{2} \sum_{i=1}^d \lambda_i\left(Z_i-1\right)\,,
    \end{aligned}
    \label{eq:analytical_S_distribution}
\end{equation}
where $\lambda_i$ are the eigenvalues of $(\mathbb{1} \pm \Sigma_2 \Sigma_1^{-1})$, $d$ is the dimensionality of the parameter space and $Z_i$ are random samples drawn from a $\chi^2$ distribution. In what follows we will drop the superscript G to simplify the notation since it will be clear in context if we refer or not to the Gaussian approximation or the full expressions.

A few points are worth noting at this point. First, the expected value of the relative entropy in Eq.~\eqref{eq:exp_kld} depends only on the posterior $p(\Theta|\mathcal{D}_1)$ and on the likelihood $\mathcal{L}(\mathcal{D}_2|\Theta)$. If we also include the assumption of linearity between the model and its parameters and Gaussianity of likelihoods, then the relative entropy in Eq.~\eqref{eq:exp_kld_analytical} depends only on the covariance matrices of the experiments. Therefore, it is a quantity solely measuring the information gain update when performing a new experiment. The PPD in Eq.~\eqref{eq:PPD}, which induces the value of $\langle D_{\rm KL} \rangle$, is a prediction of the constraining power of the new experiments with regard to the old one ($\mathcal{D}_1$). This means that the PPD, will induce this distribution independent of the data collected in a new experiment $\mathcal{D}_2$, and henceforth of the posterior $p(\Theta|\mathcal{D}_2)$.

Second, $S$ is an asymmetric measurement due to both the asymmetry in the relative entropy given by Eq.~\eqref{eq:kld} and the asymmetry in the PPD itself.
Since it is also due to the PPD, it cannot be fully resolved by using as a distance metric the Bhattacharyya distance~\citep{Bhattacharyya}, which just like the KLD is a specific case for the Rényi entropy~\citep[see][]{Pinho:2020uzv, GOLSHANI20101486}, but unlike the KLD, is symmetrical. This asymmetry is a feature rather than a failure, as further explored by~\cite{Nicola:2018rcd}.
Due to this asymmetry in $S$, one must choose with care the order of the datasets $\mathcal{D}_1$ and $\mathcal{D}_2$. An interpretation of the Surprise in the eyes of Information entropy is that the Surprise is measuring an information when going from posterior $p(\Theta|\mathcal{D}_1)$ to the posterior distribution $p(\Theta|\mathcal{D}_2)$. In that regard, it makes sense that we chose $p(\Theta|\mathcal{D}_1)$ to be less informative, or broader, than $p(\Theta|\mathcal{D}_2)$. Another interpretation could be that $\mathcal{D}_1$ comes from an experiment previous to the one yielding $\mathcal{D}_2$, like the analysis conducted in \cite{Seehars_2014, Seehars_2016}.

Finally, the choice of the KLD as the comparison term is not unique. The concept of using the PPD to induce a distribution of the metric-like quantity can be used in multiple forms. In \cite{Schosser:2024vbj} the same concept was applied in data-space using the Fisher matrix for instance.
Another possibility would be to use the Bayes factor, or the $R$ statistic, as used in the DES Year 1 analysis \citep{DES:2017myr} together with the Jeffreys' scale, and further developed in~\cite{Handley:2019wlz}. This analysis however, makes it necessary for a joint analysis of datasets to be performed, which can be problematic when dealing with strongly correlated data. This problem can be avoided by applying the Surprise~\cite{Seehars_2014} to perform a concordance analysis between WMAP and Planck data. Moreover, the relative entropy has a clear interpretation in information theory as the updated information one gain from one experiment to another~\citep{Lindley1956}. This measures both the shift in precision and disagreement between different probes. The PPD, on the other hand, measures the expected value of the relative entropy, which should also be interpreted as the expected shift in precision of the likelihoods. In that sense, the Surprise statistic can be interpreted as the information gain between probes and a measurement of disagreement between them. Finally, we emphasise that both Surprise and KLD are very robust measures, and in particular work in cases where Gaussian measures fail. For instance, there is a well-defined and finite KLD for Cauchy distributions.

We note that the Surprise can in principle be used in different regimes, depending on the choice for $\mathcal{D}_2$. If it consists of a new measurement which is independent from $\mathcal{D}_1$, we may refer to it as the replace regime. An alternative regime would be to use as $\mathcal{D}_2$ a combination of new data with the previous $\mathcal{D}_1$ data, which is often the case when aiming for higher precision. In this paper we will consider the case in which the Surprise in the replace regime. This choice was made considering that we seek to maximize our sensitivity to disagreements between datasets. An analysis of datasets which are not independent can potentially dilute tensions between the individual datasets.

\medskip
\subsection{Going beyond Gaussianity}

In this section, we evaluate the Surprise distribution using a fully numerical code and deviating from the GLM. The \texttt{klsurprise} code works by sampling the posterior predictive distribution as in Eq.~\eqref{eq:PPD_numerical}, reconstructing a posterior distribution $p(\Theta|\mathcal{D}_2^i)$ for each PPD sample $\mathcal{D}_2^i$ using Nested Sampling \citep{Skilling:2004pqw, Skilling:2006gxv}, and then performing the Monte Carlo integral in Eq.~\eqref{eq:kld_numerical}. The resulting vector will be used to reconstruct the distribution of $D_{\rm KL}$, its mean value and consequently, the distribution of the Surprise, $S$. This distribution can then be used with the value of the numerical Surprise statistic to compute a $p$-value.  Appendix~\ref{app:code} describes the \texttt{klsurprise} code in detail.

It is important to note that although time-consuming, for parameter spaces with few dimensions the \texttt{klsurprise} code can be run on standard modern desktops. In fact, for the case with $\mathcal{D}_1 = $ BOSS+$\,$\&$\,$BBN and $\mathcal{D}_2 = $ Pantheon+$\,$\&$\,$SH0ES in o$w$CDM, which has 4 free parameters, the total computation time was a modest 540 CPU hours. This suggests that one does not need to rely on the Gaussian approximation on the grounds of computational cost alone, at least when the number of parameters is small.

\subsection{Gaussian vs.~non-Gaussian $S$}

In Bayesian analysis, projections effects can create a false sense of agreement by failing to represent a complex $n$-dimensional distribution in a one or two-dimensional plane. As noted and explored by \citep{Raveri:2021wfz, Handley:2019wlz, DES:2020hen, Lin:2017ikq} in the GLM framework, these projections can hide inconsistencies between datasets. In that sense, the Surprise is a useful tool to identify such inconsistencies that might be hidden by marginalization. Consider, for instance, the illustrative example of two SNIa mocks in a o$w$CDM model. Both were generated with a diagonal covariance matrix with a distance modulus error of $\sigma_\mu = 0.14$. The values of $h$ are fixed in the likelihood analysis, and the fiducial parameters of the mocks are given in Table~\ref{tab:fiducial}.

\begin{figure}[t]
    \centering
    \includegraphics[width=0.9\linewidth]{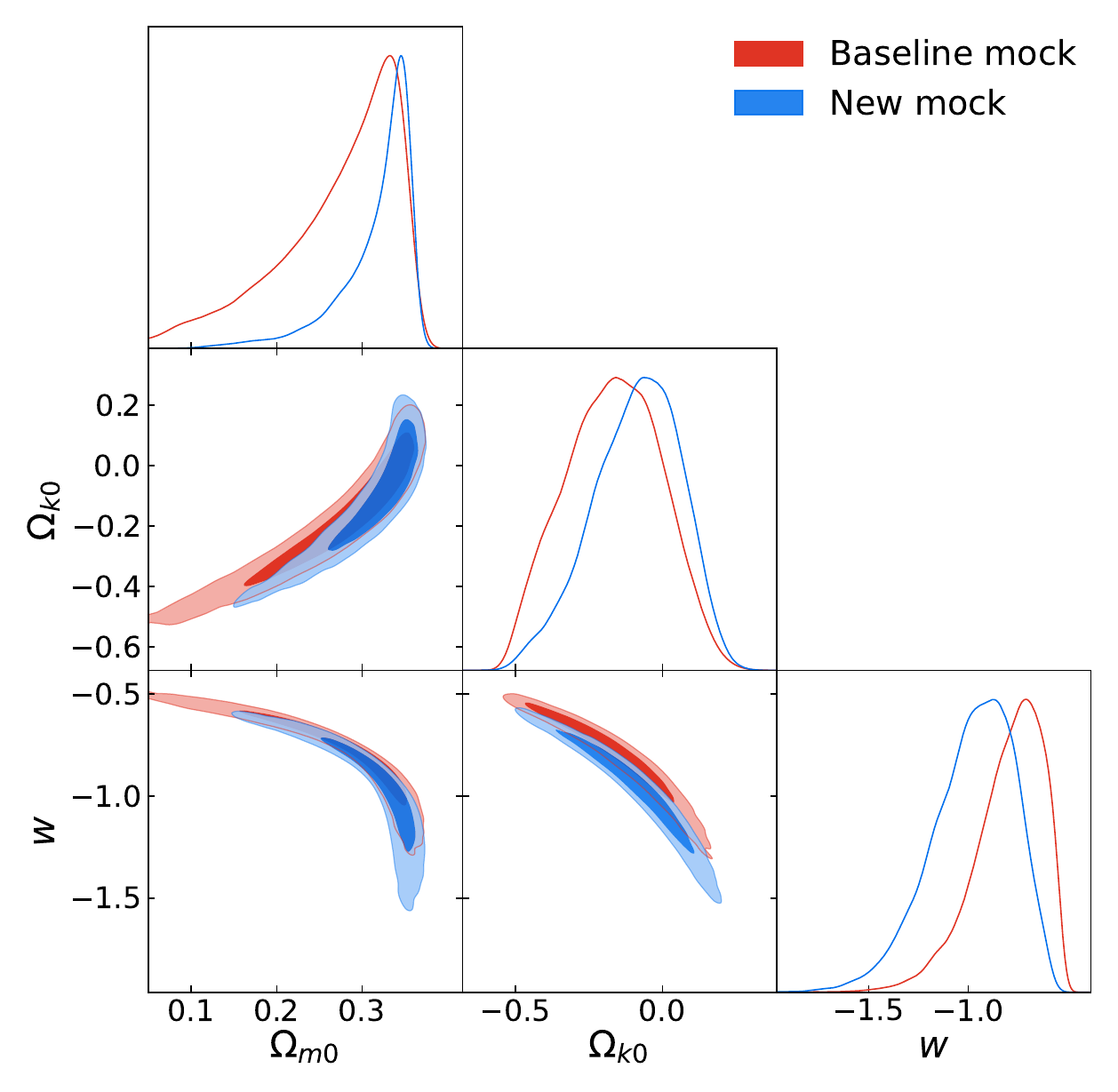}
    \caption{Marginalized 1 and 2$\sigma$ (i.e.~68.3\% and 95.4\% CI) contour plots for our supernova mocks.}
    \label{fig:marginalize_contours_SNIa}
    \medskip
\end{figure}

\begin{figure}[t]
    \centering
    \includegraphics[width=0.8\linewidth]{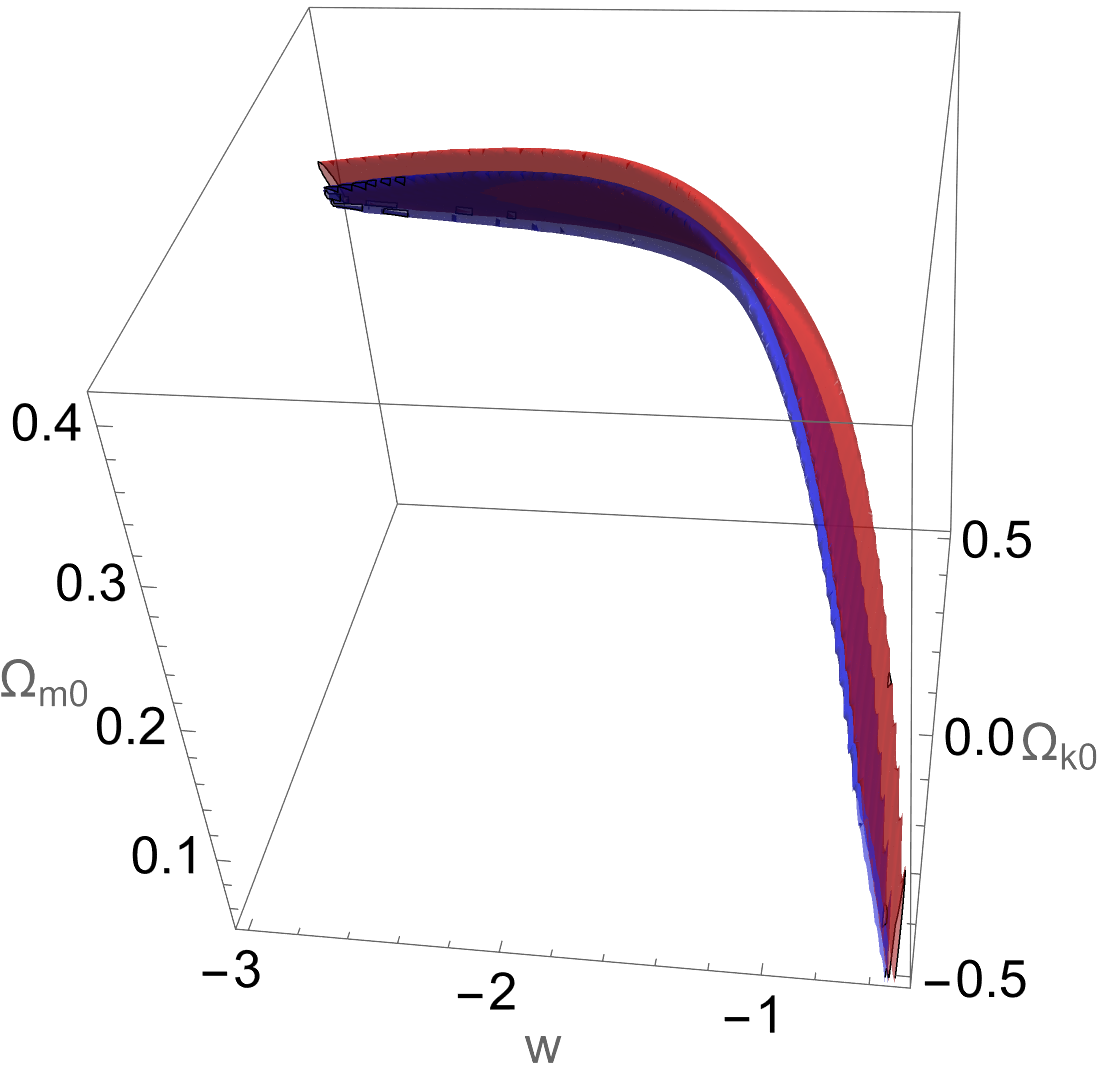}
    \includegraphics[width=0.85\linewidth]{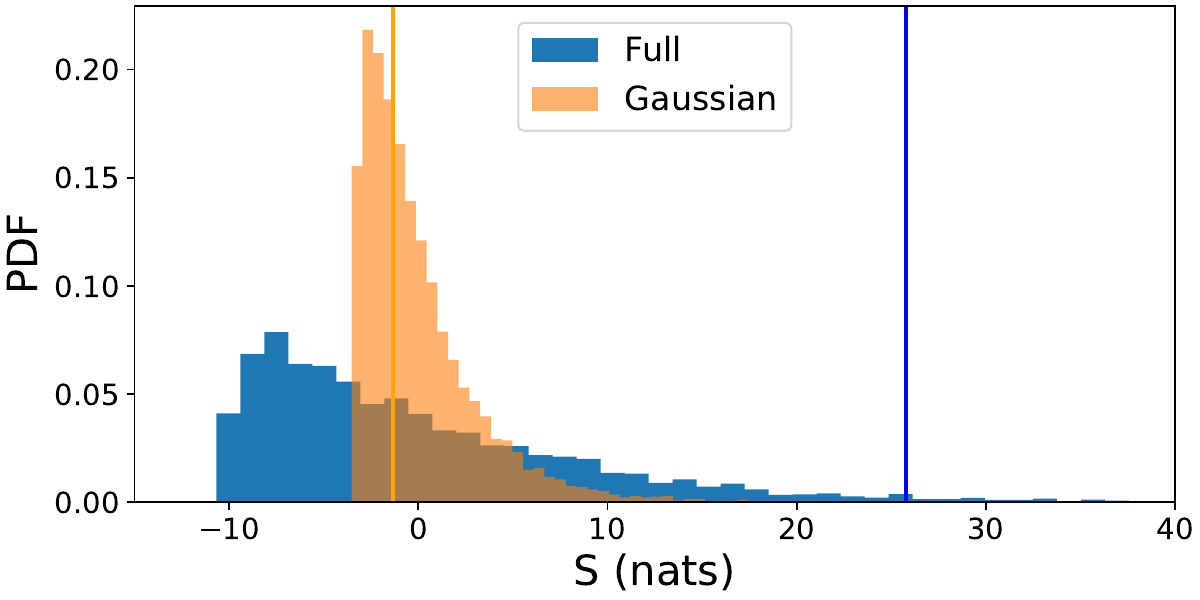}
    \caption{
    \emph{Top:} three-dimensional 95\% confidence contour plots for SNIa mocks. \emph{Bottom:} surprise statistic distribution for the mocks in Table~\ref{tab:fiducial}.}
    \label{fig:SNIa_mocks_full}
    \bigskip
\end{figure}

The contour plots of the posterior distribution for these mocks are shown in Figure~\ref{fig:marginalize_contours_SNIa} and by themselves show no apparent large discordance between the datasets, as 2$\sigma$ contours intersect.  This consistency, though, is only apparent, as the banana shaped contours of SN barely overlap when the contour plots are made in 3 dimensions, as shown in Figure~\ref{fig:SNIa_mocks_full}. These hidden discordances can appear in multidimensional data and remain undetected when looking only at marginalized contours. The Surprise can account for these discrepancies, even for high dimensional data by translating the effective complexity of distributions into a one-dimensional distribution for S and also consequently a $p$-value statistics, which will indicate the probability for measuring a Surprise value that deviates from zero by more than S. However, approximating the posterior distributions by Gaussians can sometimes make two disagreeing distributions falsely agree, as Gaussian approximations cannot fully describe the shape of some probability density functions.

\begin{table}[t]
    \centering
    \begin{tabular}{ccccc}
        \hline\hline
          & $h$   & $\Omega_{m0}$ & $\Omega_{k0}$ & $w$\\ \hline
        Baseline mock & 0.7 & 0.35       & 0.0        & -1\\
        New mock  & 0.7 & 0.34       & -0.03       & -1.03\\
        \hline\hline
    \end{tabular}
    \caption{Fiducial parameters of our mock supernova data.}
    \label{tab:fiducial}
\end{table}

For the specific case of the mocks constructed here, one can see that the Surprise deviate significantly from its linear Gaussian solution. The distribution of the Surprise is shown in Figure~\ref{fig:SNIa_mocks_full}. The Gaussian Surprise distribution was generated by sampling $\chi^2$ distributions in Eq.~\eqref{eq:analytical_S_distribution}.
The Gaussian Surprise distribution presents a much tighter distribution than that of the numerical Surprise. The relevant statistical quantities for this particular case can be seen in Table~\ref{tab:surprise_results_mock_SNIa}.  Using the global covariance of the posterior to approximate them as Gaussians, effectively remove any discordances that could be fully appreciated in the full Surprise statistic. As can be seen from Table~\ref{tab:surprise_results_mock_SNIa}, the full result indicates a clear tension in datasets, while the analytical solution for the Surprise, which uses the GLM, does not.

\medskip
\subsection{$p$-values and $\sigma$-values}

Throughout this manuscript, we often convert $p$-values to a significance in Gaussian $\sigma$ (sometimes called $\sigma$-values) for simple convenience, even when not assuming Gaussian posteriors. In other words, from the $p$-values we calculate the $\sigma$-values with
\begin{equation}\label{eq:sigma}
    \sigma{\text{-value}} = \sqrt{2} \; {\rm erf}^{-1}(1 - p{\text{-value}}).
\end{equation}

%%%%%%%%%%%%%%%%%%%%%%%%%%%%%%%%%%%%
\bigskip
\smallskip
\section{Data and Likelihoods}
\label{sec:data}
%%%%%%%%%%%%%%%%%%%%%%%%%%%%%%%%%%%%

In this section, we present and discuss the datasets employed in our cosmological analysis using the non-Gaussian Surprise. Specifically, we examine the Pantheon+$\,$\&$\,$SH0ES SNIa dataset~\citep{Riess:2021jrx,Brout:2022vxf} and the peak locations of Baryonic Acoustic Oscillations (BAO).

\begin{table}[t]
    \centering
    \begin{tabular}{lcc}
    \hline\hline
            & \textbf{Gaussian} & \textbf{Full} \\
    \hline
    $D_{\rm KL}$                   & 2.80       & 36.4       \\
    $\langle D_{\rm KL} \rangle$ & 4.03       & 10.6      \\
    $S$                   & -1.23      & 25.8      \\
    $p$-value               & 58\%     & 2.6\%     \\
    significance ($\sigma$) & 0.6 & 2.3 \\
    \hline\hline
    \end{tabular}
    \caption{Values for the Surprise statistic between two SNIa mocks.}
    \label{tab:surprise_results_mock_SNIa}
    \medskip
\end{table}

\begin{table*}[t]
\centering
\begin{tabular}{lcccc}
    \hline\hline
     %&   & \multicolumn{3}{|c|}{\textbf{Value} }\\    \cline{3-5}
     & $z_\text{eff}$ & \textbf{$D_{\rm M}/r_{\rm d}$} & \textbf{$D_{\rm H}/r_{\rm d}$} & $D_{\rm V}/r_{\rm d} \;{\rm or}\; r$ \\
    \hline
    \multicolumn{5}{c}{\textbf{BOSS+}}\\ \hline
    6DF                           & 0.106           &---&---& $r_{\rm d}/D_{\rm V} = 0.327 \pm 0.015$\\
    SDSS DR7 MGS                  & 0.15            &---&---& $4.47 \pm 0.16$\\
    SDSS DR14 eBOSS QSO           & 1.52            &---&---& $26.0 \pm 1.0$\\
    SDSS DR12 BOSS galaxies            & 0.38, 0.51, 0.61 & full covariance & full covariance &---  \\
    SDSS DR14 eBOSS Ly$\alpha$ & 2.34   &$37.41\pm 1.87$& $8.86 \pm 0.29$&---  \\
    \hline
    \multicolumn{5}{c}{\textbf{DESI}}\\ \hline
    BGS     & 0.30 & --- & --- & $7.93 \pm 0.15$ \\
    LRG     & 0.51 & $13.62 \pm 0.25$ & $20.98 \pm 0.61$ & $r=-0.445$ \\
    LRG     & 0.71 & $16.85 \pm 0.32$ & $19.90 \pm 0.51$ & $r=-0.420$ \\
    LRG+ELG & 0.93 & $21.71 \pm 0.28$ & $17.88 \pm 0.45$ & $r=-0.389$ \\
    ELG     & 1.32 & $27.79 \pm 0.69$ & $13.82 \pm 0.42$ & $r=-0.444$ \\
    QSO     & 1.49 & --- & --- & $26.07 \pm 0.67$ \\
    Ly$\alpha$ QSO  & 2.33 & $39.71 \pm 0.94$ & $8.52 \pm 0.17$ & $r=-0.477$ \\
    %\hline
    \hline
    \hline
    \end{tabular}
    \caption{Comprehensive list of the BAO data used.
    \emph{Top:} the BOSS+ combination: 6dF + SDSS BOSS + eBOSS.
    \emph{Bottom:} DESI. Note that for each sample we measure either both $D_{\rm M}/r_{\rm d}$ and $D_{\rm H}/r_{\rm d}$, which are correlated with a coefficient $r$, or $D_{\rm V}/r_{\rm d}$.
    \label{tab:BAO_data}}
    \medskip
\end{table*}

\medskip
\subsection{Pantheon+ Likelihood}

The SNIa dataset employed in this study is sourced from Pantheon+ catalogue. These supernovae span a redshift range from \( z = 0.001 \) to \( z = 2.26 \). The computation of the distance modulus incorporates calibrations based on Cepheid variable distances from the SH0ES collaboration \cite{Riess:2021jrx}. The distance modulus is related to the luminosity distance as
\begin{equation}
    \mu(\Theta| z) = 5\log_{10} d_L(\Theta| z) + 25,
    \label{eq:distance_modulus}
\end{equation}
the luminosity distance $d_L$ itself is a function of redshift and cosmological parameters \citep{Hogg:1999ad}.

The likelihood is built under the assumption of Gaussianity:
\begin{equation}
    \mathcal{L}(\boldsymbol{\mu}|\Theta) = \frac{1}{\sqrt{(2\pi)^d |\boldsymbol{\Sigma}|}} \exp\left(-\frac{1}{2} \Delta\boldsymbol{\mu}^\top \boldsymbol{\Sigma}^{-1} \Delta\boldsymbol{\mu}\right),
    \label{eq:SNIa_likelihood}
\end{equation}
where $d$ is the dimensionality of the dataset $\Delta\mu_i = (\mu_i^{\rm obs} - \mu_i(\Theta|z))$ and $\mu_i(\Theta|z_i)$ is the theoretical prediction for the distance modulus of the $i^{\rm th}$ SNIa in the catalogue.

\medskip
\subsection{BAO+BBN Likelihood}

For the analysis of BAO we utilized data from the BOSS+ combination. A comprehensive table of the data used for the BOSS+ likelihood is given in Table~\ref{tab:BAO_data}. The other likelihood utilized is from the DESI 2024 data release~\citep{DESI:2024mwx}, for which the dataset is given in Table~\ref{tab:BAO_data}. These datasets measure different distances at different redshifts. They are not used together, as they possess intercepting redshift bins, and therefore, non-zero correlations. To build the likelihood we use only the BAO peak locations, i.e., the redshift-dependent BAO scale $(1/Hr_{\rm s})$, the angular BAO scale $(D_{\rm A}/r_{\rm s})$ and the volume averaged BAO scale $(D_{\rm V}/r_{\rm s})$.  The DESI data has  six different traces at different redshifts. The bright galaxy sample (BGS) are low redshift galaxies with $z$ ranging from 0.1 to 0.4, the luminous red galaxy sample (LRG), the emission line galaxies (ELG) and Lyman$-\alpha$  (Ly$\alpha$) tracers. In our analysis, a Gaussian likelihood is assumed for the data-vector and the cosmological background quantities, namely $D_{\rm A}$, $D_{\rm V}$, and $H$, are calculated using the Python package jax-cosmo~\citep{Campagne_2023}.

Given that the BAO likelihood exhibits degeneracy in the $H_0 - r_{\rm s}$ plane, we employ Big Bang Nucleosynthesis (BBN) data as a prior for $\omega_b$ to estimate the sound horizon radius at the drag epoch, $r_{\rm s}$. Although a Boltzmann code such as CLASS \citep{class} is typically required for precise computation of $r_{\rm s}$, we opted for a numerical approximation as in \cite{BOSS:2014hhw}:
\begin{equation}
        r_{\rm s}(z_d) \approx \frac{55.154 \exp\left[-72.3(\omega_b + 0.0006)^2\right]}{\Omega_{m0}^{0.25351}\omega_b^{0.12807}} \, \text{Mpc},
        \label{eq:rs_approximation}
\end{equation}
where $\omega_X = \Omega_X h^2$, with $X = m, \nu, b$ for matter, neutrinos and baryons, and $H_0=100\,h\, \mathrm{km/s/Mpc}$.

The Big Bang Nucleosynthesis (BBN) dataset leverages the abundance of deuterium to infer the density of baryons, given the correlation between these two quantities. Due to the absence of significant post-Big Bang production mechanisms for deuterium, its primordial abundance is deduced from observations of systems with low metallicities (see \cite{Cyburt:2015mya} for a review). These observations allow for the derivation of the baryon-to-photon ratio, $\eta$. Thus, observational constrains allow for the determination of primordial deuterium abundance and helium fraction, which can be used to deduce $\omega_b  = \Omega_b h^2$. The constraints used for $\Omega_b h^2$ depend on the theoretical treatment of nuclear interaction cross-sections, mostly for the deuterium burning reactions. In this work, we use the predictions given in \cite{Schoneberg:2024ifp}, which make use of the code {\bf PRyMordial} \citep{Burns:2023sgx} and reports the constraints:
\begin{equation}
    \omega_b = \Omega_b h^2 = 0.02218 \pm 0.00055.
    \label{eq:bbn_prior}
\end{equation}

%%%%%%%%%%%%%%%%%%%%%%%%%%%%%%%%%%%%%%%%%%%%%%%%%%%%%%%%%%%%%%%%%%%%%%%%%%%%%%%%%%%%%%%%%%%%%%%%%%%%%%%%%%%%%%%%%

%%%%%%%%%%%%%%%%%%%%%%%%%%%%%%%%%%%%
\bigskip
\section{Non-Gaussian Surprise in Cosmology}\label{sec:analysis}
%%%%%%%%%%%%%%%%%%%%%%%%%%%%%%%%%%%%

As mentioned previously, we will adopt as our reference model the o$w$CDM model. In most models beyond $\Lambda$CDM, the dark energy equation of state parameter $w$ is a time varying parameter. This is often taken into account using the simple CPL parametrization~\citep{Linder:2002et, Chevallier:2000qy}, which adds a $w_a$ parameter to account for this time-evolution of $w$. In many current cosmological probes, however, $w_a$ is poorly constrained by the data, and thus the simpler model $w$CDM, with a fixed $w$, is used instead.  Recent surveys such as DESI~\citep{DESI:2024mwx} have shown, when combined with supernova data, hints of a non-zero $w_a$, putting a new tension in the standard $\Lambda$CDM scenario, although results are still heavily influenced by the choice of prior~\citep{Cortes:2024lgw}. Since here our focus is to analyse the Surprise  between pairs of probes without combining them, we choose to focus on the o$w$CDM, without including $w_a$.

\begin{figure}[t]
    \centering
    \centering
    \includegraphics[width=\columnwidth]{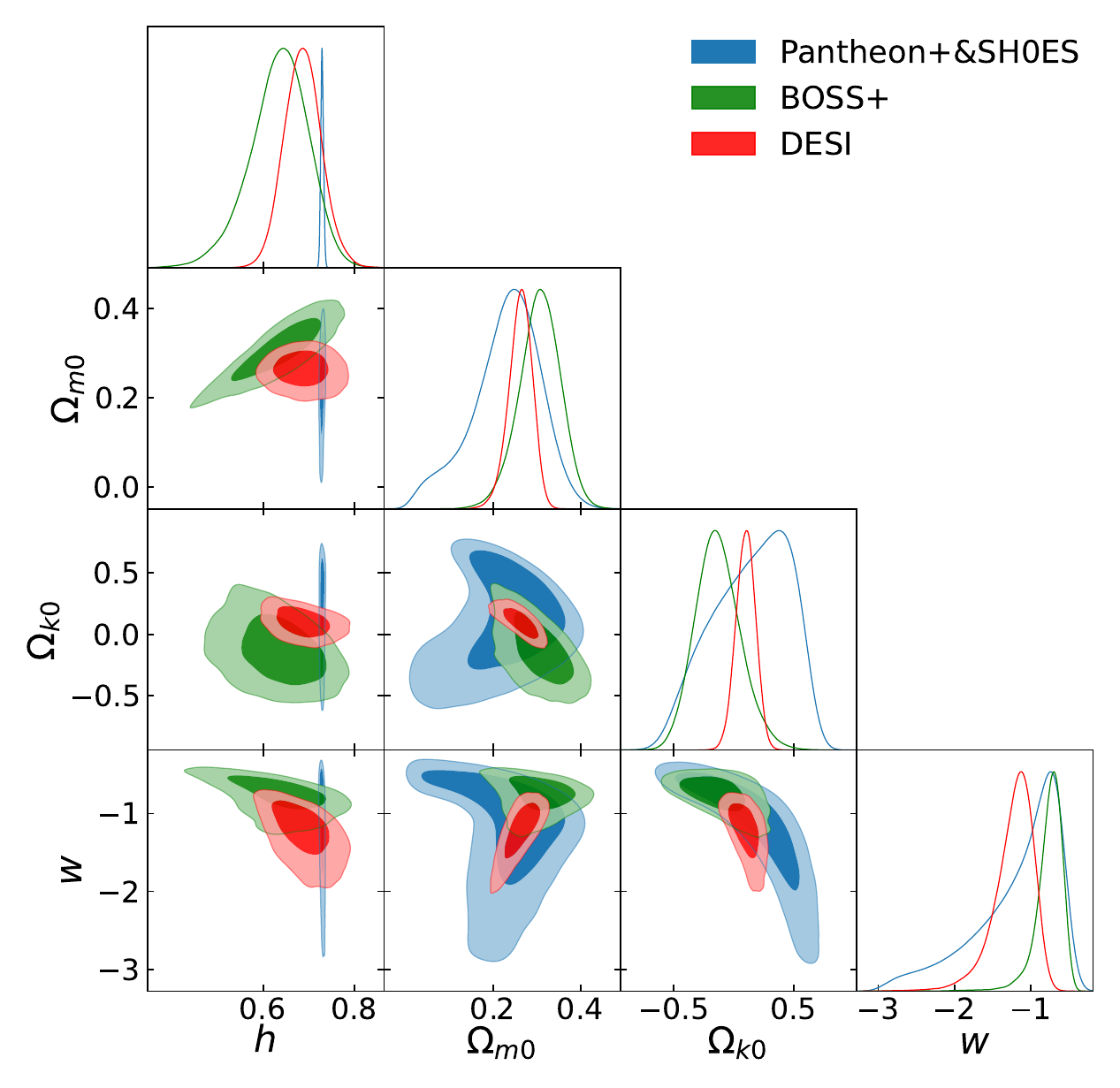}
    \caption{
    1 and 2$\sigma$ contours for the posteriors of DESI (red), BOSS+ (green) and Pantheon+ (blue) considering an o$w$CDM cosmological model.
    }
    \label{fig:DESI_BOSS_Pantheon}
    \medskip
\end{figure}

\begin{figure}[t]
    \centering
    \includegraphics[width=.9\columnwidth]{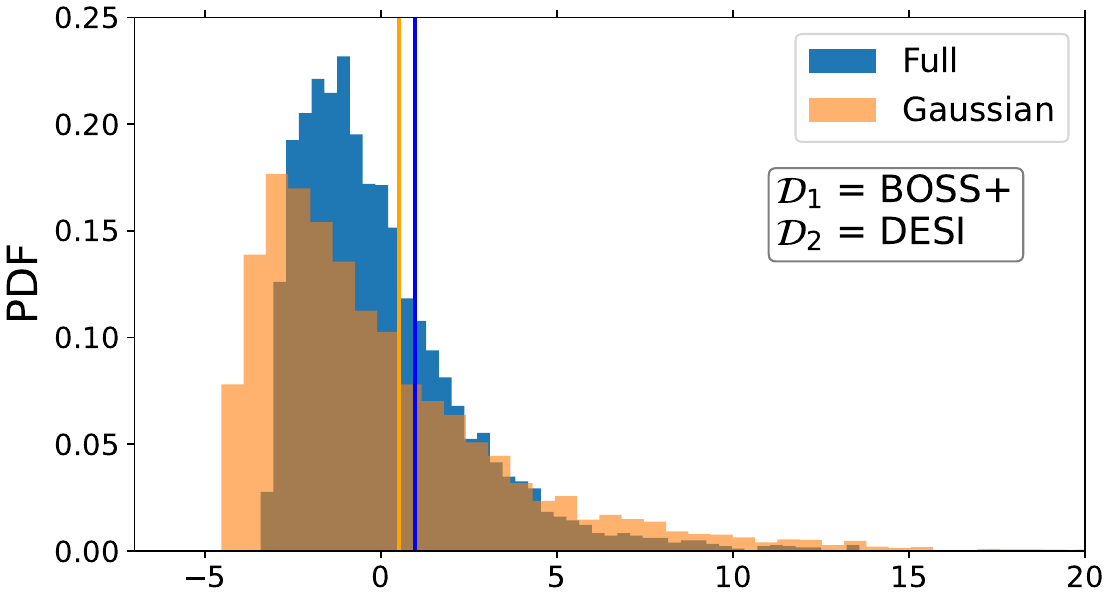}
    \includegraphics[width=.9\columnwidth]{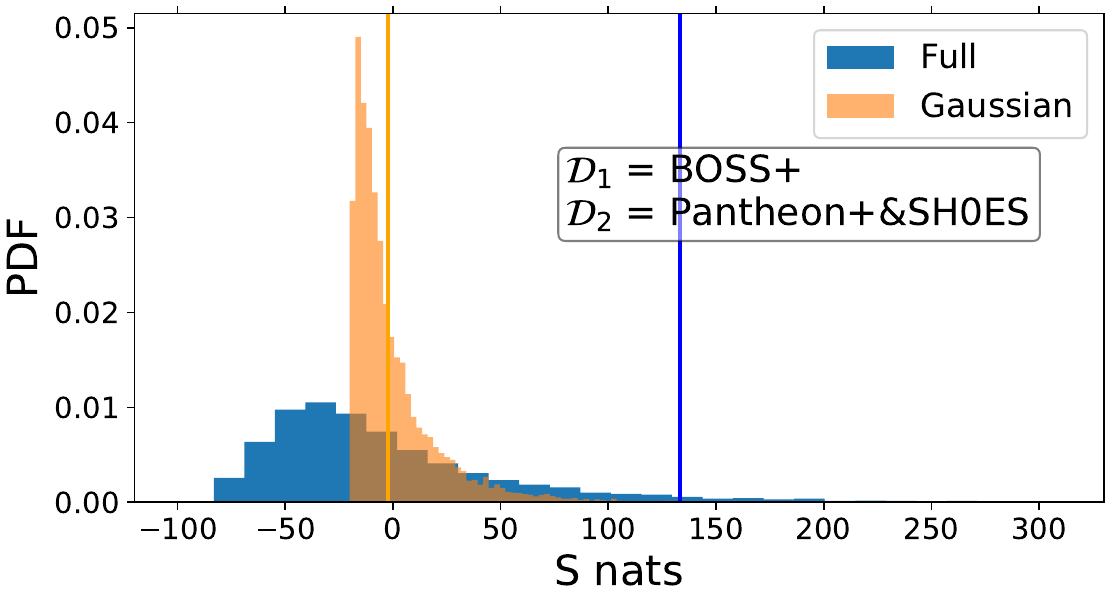}
    \includegraphics[width=.9\columnwidth]{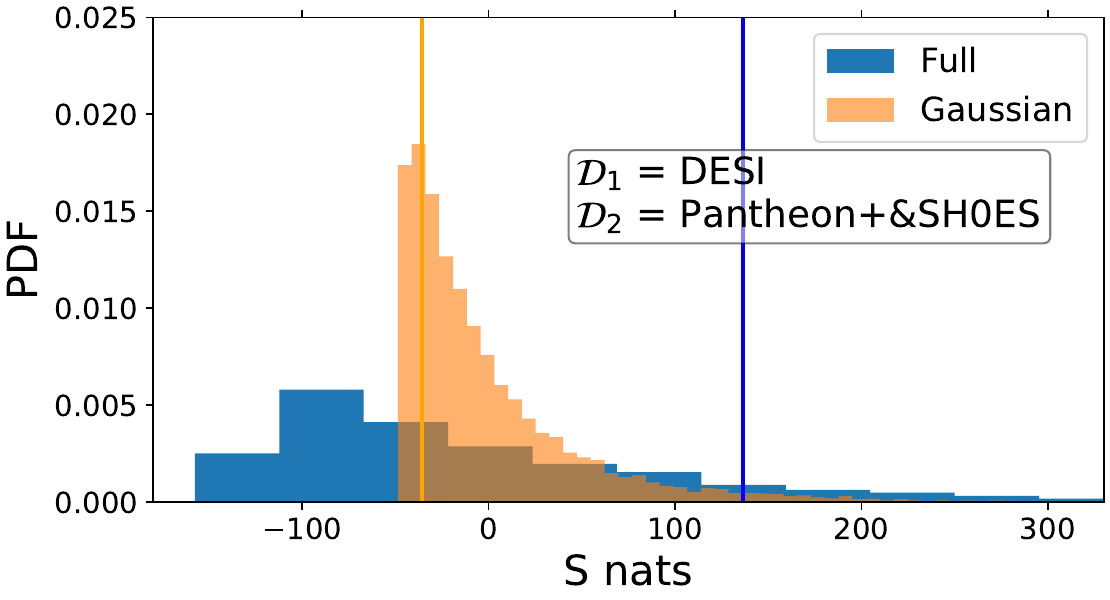}
    \caption{
    Surprise distribution for the o$w$CDM model, for the analytical Gaussian approximation (orange) and full numerical solution (blue). The vertical lines represent the actual Surprise values.
    \emph{Top:} DESI as $\mathcal{D}_2$, BOSS+ as $\mathcal{D}_1$.
    \emph{Middle:} Pantheon+ as $\mathcal{D}_2$, BOSS+ as $\mathcal{D}_1$.
    \emph{Bottom:} Pantheon+ as $\mathcal{D}_2$, DESI as $\mathcal{D}_1$.
    We note tensions between Pantheon+$\,$\&$\,$SH0ES and DESI/BOSS+ probes (see also Table~\ref{tab:BAO_results}).The Gaussian $S$ is not a good approximation when using supernovae, but not too bad for BAO data alone.
    } \medskip
    \label{fig:S-histograms}
\end{figure}

Figure~\ref{fig:DESI_BOSS_Pantheon} shows the 1 and 2$\sigma$ contours in the o$w$CDM model for the main datasets here considered: BOSS+, DESI and Pantheon+$\,$\&$\,$SH0ES. On inspection  of this figure, DESI seems reasonably Gaussian, whereas some non-Gaussianity is present in BOSS+ and considerable non-Gaussianities are apparent on Pantheon+$\,$\&$\,$SH0ES. This hints at the fact that while a Surprise analysis of purely BAO data assuming Gaussianity may lead to reasonable results, if one involves the supernova data $S$ must be computed using the full numerical treatment.

Figure~\ref{fig:S-histograms} illustrates the different distributions for $S$ assuming the o$w$CDM model, for both the full numerical solution and the Gaussian approximation. We show the three possible combinations of datasets. One can see that, as expected, the surprise between both BAO datasets can be reasonably approximated by its Gaussian solution. However, large deviations from the simple Gaussian solution when the supernova data is considered. We will discuss this in more detail below for each pair of data.

\medskip
\subsection{DESI vs.~BOSS+}

\begin{figure}[t]
    \centering
    \includegraphics[width=0.9\linewidth]{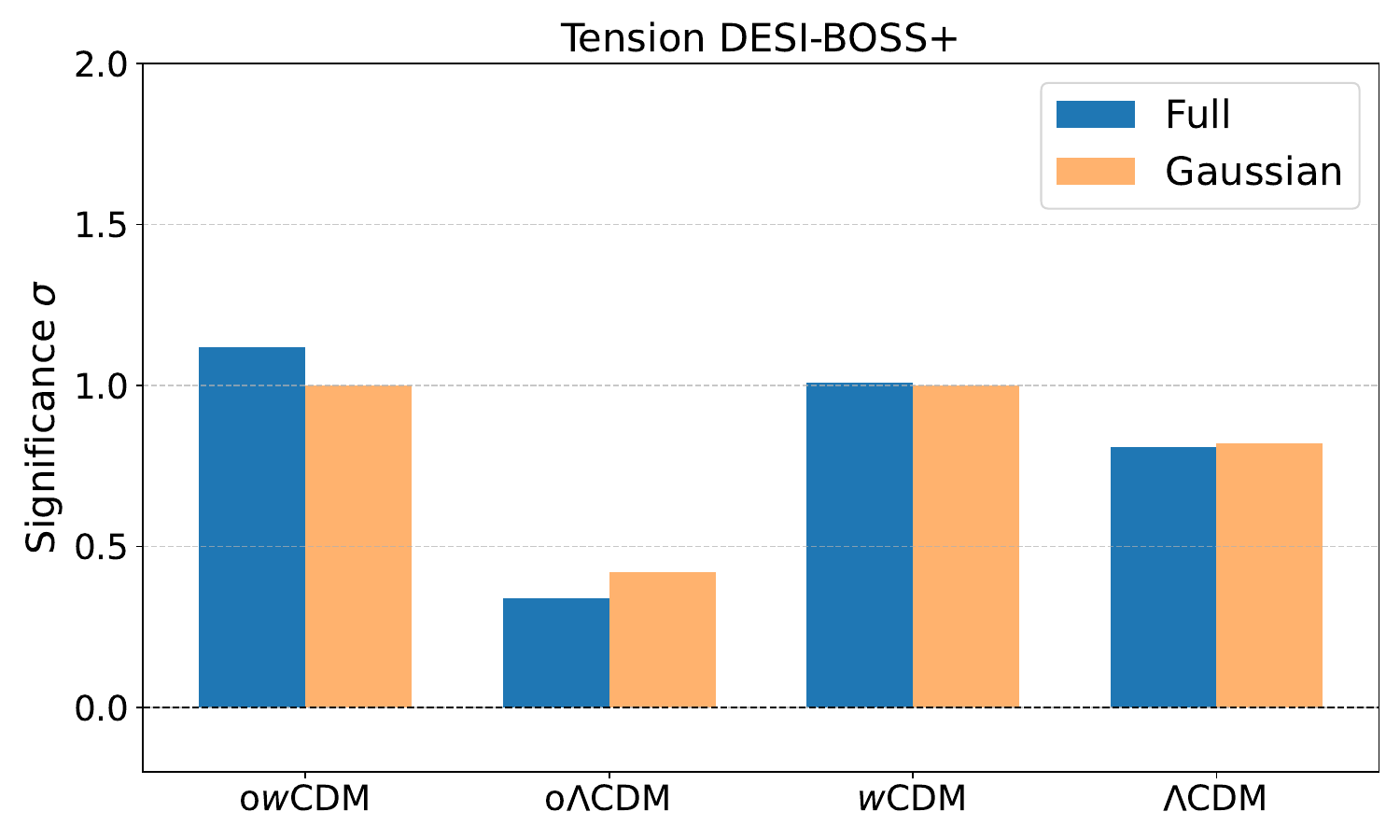}
    \caption{ Tension in $\sigma$ considering $\mathcal{D}_1 = $ BOSS+ and $\mathcal{D}_2 = $ DESI. The tension was evaluated by computing the significance of the $p$-value from the Surprise statistic and its distribution for different cosmological models.}
    \label{fig:tension-DESIBOSS}
    \medskip
\end{figure}

One of the main advantages of the Surprise is that it can be used with correlated experiments without the need to perform a joint analysis. In that sense, an obvious use of the Surprise is to compare DESI and BOSS+ experiments, which measure intersecting volumes therefore have non-zero correlations. In this particular case, the asymmetry of the Surprise distribution can be easily interpreted, chronologically BOSS+ is a previous experiment to DESI. I.e., it's natural to use $\mathcal{D}_1$ = BOSS+ and $\mathcal{D}_2$ = DESI.  For these two datasets in a o$w$CDM model, both the analytical and the numerical results for the Surprise indicate no apparent tension, as can be seen from results in Table~\ref{tab:DESI_SDSS_Pantheon_DESI_results}. The validity of the Gaussian $S$ can be seen in the top panel of Figure~\ref{fig:S-histograms} and also in the results of Table~\ref{tab:DESI_SDSS_Pantheon_DESI_results}.
The tension computed from the Surprise  barely surpasses the $1\sigma$ value, indicating a good cosmological agreement between BOSS+ and DESI. A summary of the tension (in)significance between BOSS+ and DESI is depicted in Figure~\ref{fig:tension-DESIBOSS} for the different cosmological models here considered.

\medskip
\subsection{The tension between Pantheon+ and BAO+BBN.}

\begin{table}[t]
\begin{tabular}{lcc|cc}
    \hline\hline
                                              & \multicolumn{2}{l}{DESI|BOSS+} & \multicolumn{2}{l}{Pantheon+|DESI} \\ \hline
                                              & \textbf{Gaussian} & \textbf{Full} & \textbf{Gaussian}   & \textbf{Full}   \\ \hline
$D_{\rm KL}$                                    & 5.25              & 5.61          & 62.0                & 51.9            \\
$\langle D_{\rm KL} \rangle$ & 5.21              & 4.28          & 102.8                & 46.6            \\
$S$                                  & 0.04              & 1.33          & -40.8               & 5.36            \\
$p$-value                                       & 35\%            & 22\%        & 78\%              & 26\%          \\
significance ($\sigma$)                       & 0.9             & 1.2          & 0.3                & 1.1           \\
\hline \hline
\end{tabular}
    \caption{Values for the Surprise statistic $S$(DESI|BOSS+) and $S$(Pantheon+|DESI), both in a o$w$CDM cosmological model.
    }
    \label{tab:DESI_SDSS_Pantheon_DESI_results}
\end{table}

We now consider the Surprise between the supernova and BAO+BBN datasets. We will  consider $\mathcal{D}_1$ = DESI or BOSS+ and $\mathcal{D}_2$ = Pantheon+$\,$\&$\,$SH0ES. The BBN prior of $\omega_b$ is applied and marginalized over in the BAO+BBN likelihood, so the parameter space is restricted to a common set of parameters for both experiments.

Table~\ref{tab:BAO_results} below summarizes the surprise for all four cosmological models here considered (o$w$CDM, oCDM, $w$CDM, $\Lambda$CDM), and for both the full numerical $S$ and its Gaussian approximation. In what follows, we will discuss the results for each cosmological model separately.

\subsubsection{$S$ for the {\rm o$w$CDM} model}

\begin{table*}[t]
    \centering
    \begin{tabular}{ccccccccc}
    \hline\hline
    & \multicolumn{2}{c}{{o$w$CDM}} & \multicolumn{2}{c}{{o$\Lambda$CDM}} & \multicolumn{2}{c}{{$w$CDM}} & \multicolumn{2}{c}{{$\Lambda$CDM}}  \\ \hline
    & {Gaussian} & {Full} & {Gaussian} & {Full} & {Gaussian} & {Full} & {Gaussian} & {Full} \\
    \hline
    \multicolumn{9}{c}{\textbf{BOSS+ vs. Pantheon+$\,$\&$\,$SH0ES}}\\
    \hline
    $D_{\rm KL}$ (nats) & 33.8 & 246 & 65.5 & 87.1 & 91 & 181 & 7.1 & 7.1 \\
    \textbf{$\langle D_{\rm KL} \rangle$} (nats) & 36.4 & 112.8 & 34.6 & 44.8 & 58.7 & 111.9 & 3.8 & 3.9 \\
    S (nats) & -2.6 & 133 & 30.9 & 42.3 & 32 & 69 & 3.2  & 3.2 \\
    $p$-value & 38\% & 4.5\% & 8.8\% & 10\% & 14\% & 17\% & 5.6\% & 5.2\% \\
    {significance ($\sigma$)} & 0.9 & 2.0 & 1.7 & 1.6 & 1.5 & 1.4 & 1.9 & 1.9 \\ \hline
    \multicolumn{9}{c}{\textbf{DESI vs. Pantheon+$\,$\&$\,$SH0ES}}\\
    \hline
    $D_{\rm KL}$ (nats) & 56.1 & 334 & 110 & 135 & 87 & 229 & 19.1 & 18.5 \\
    \textbf{$\langle D_{\rm KL} \rangle$} (nats) & 92 & 197.5 & 55.5 & 71.1 & 56.3 & 71.2 & 3.6 & 3.3 \\
    S (nats)  & -35.9 & 136 & 54.9 & 64.4 & 31.0 & 158 & 15.5 & 15.1 \\
    $p$-value & 76\% & 12\% & 8.0\% & 11\% & 15\% & 5.9\% & 0.10\% & 0.08\% \\
    {significance ($\sigma$)} & 0.3 & 1.5 & 1.8 & 1.6 & 1.4 & 1.9 & \high{3.3} & \high{3.4} \\ \hline  \hline
    \end{tabular}
    \caption{Results for the $S(p(\Theta|\mathcal{D}_2)\mid p(\Theta|\mathcal{D}_1))$ for both Gaussian approximation and the full numerical result, for different cosmological models. Here $\mathcal{D}_1$ is either BOSS+ or DESI (BAO+BBN) and $\mathcal{D}_2$ is Pantheon+$\,$\&$\,$SH0ES. In all tables we mark in red values with significance above $2\sigma$.}
    \label{tab:BAO_results}
    \smallskip
\end{table*}

We analyse first the full o$w$CDM model, since it is the one with the most degrees of freedom. The o$w$CDM model also offers the best insight on the behaviour of the non-Gaussian Surprise, as the data is a highly non-linear function of the parameters and the contour plots deviate significantly from Gaussianity, as can be seen in Figure~\ref{fig:DESI_BOSS_Pantheon}. In fact, again as expected, the Surprise distribution deviates significantly from the solution assuming the Gaussian linear model, as can be seen in the bottom panel of Figure~\ref{fig:S-histograms}.

\begin{figure}[t]
    \centering
    \includegraphics[width=0.8\linewidth]{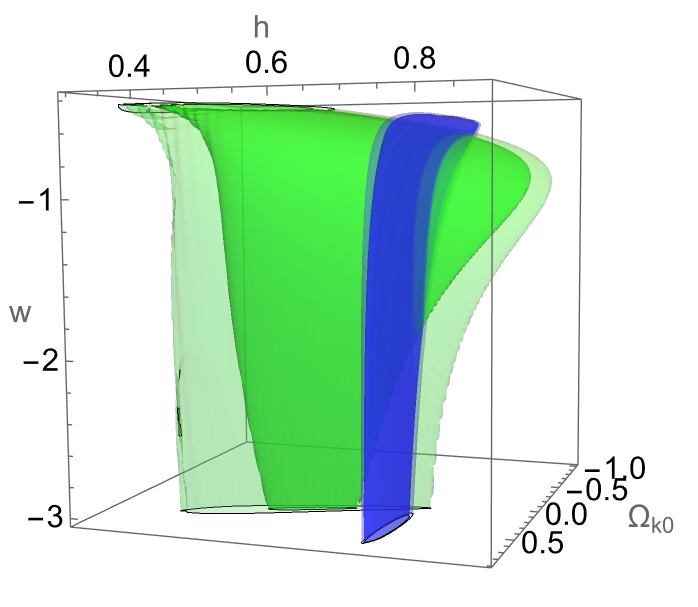}
    \includegraphics[width=0.8\linewidth]{figs/plot_3D_DESI_Pantheon_owCDM_margOm.png}
    \caption{3D posteriors contour plots of 68\% and 95\% CI.
    \emph{Top:} Pantheon+$\,$\&$\,$SH0ES (blue) and BOSS+ posterior (green), marginalizing over $\Omega_{m0}$.
    \emph{Bottom:} Pantheon+$\,$\&$\,$SH0ES (blue) and DESI (red), marginalizing over $H_0$.    \label{fig:3D_contours}
    }
    \medskip
\end{figure}

For the case where $\mathcal{D}_1 = $ BOSS+ the full solution for the Surprise finds a $p$-value of 4.5\%, indicating tension between the datasets. This tension appears in the 4-dimensional parameter space, but most of its contribution comes from the known $H_0$ tension. We can visualize the contour plots in 3-dimensions by marginalizing over $\Omega_{m0}$ to try and understand the origins of the tensions. This is depicted in Figure~\ref{fig:3D_contours}. We see that in the $\{h, \Omega_{k0}, w\}$ space there is a very limited intersection between the posterior PDFs, causing high values of the relative entropy between both distributions and, therefore, a high value of $S$. This explains the discordance between seen in the top panel of Table~\ref{tab:BAO_results}.

When analysing the same model, but considering DESI as our $\mathcal{D}_1$, the $p$-value increases to 12\%. The Surprise distribution for this case still deviates significantly from the GLM (bottom panel of Figure~\ref{fig:S-histograms}). Both the expected relative entropy and the relative entropy have significantly different values than their counterpart solutions in the GLM.

Again, we can analyse the contour plots in the 3D $\{h, \Omega_{k0}, w\}$ space as seen in Figure~\ref{fig:3D_contours} to see where the discrepancy of DESI and Pantheon+ data comes from.  We see that the contour plots of DESI shrink significantly in relation to BOSS+. This causes the overlap between the distribution to increase in relation to the total distribution mass, which causes a decrease in the value of the Kullback-Leibler divergence and thus the Surprise. The distributions, however, present a 1.5$\sigma$ discrepancy.

In summary, BOSS+ and DESI have no significant tension, BOSS+$\,$\&$\,$BBN and Pantheon+$\,$\&$\,$SH0ES have a slight 2.0$\sigma$ tension, DESI+BBN and Pantheon+$\,$\&$\,$SH0ES have a 1.5$\sigma$ discrepancy. Marginalizing results over $h$, we find that this slight tension between DESI and Pantheon+ disappears. This means that the tension in data is, unsurprisingly, introduced by the tension in $h$.

\begin{figure*}[t!]
    \centering
    \begin{minipage}{0.9\columnwidth}
        \centering
        \includegraphics[width=\linewidth]{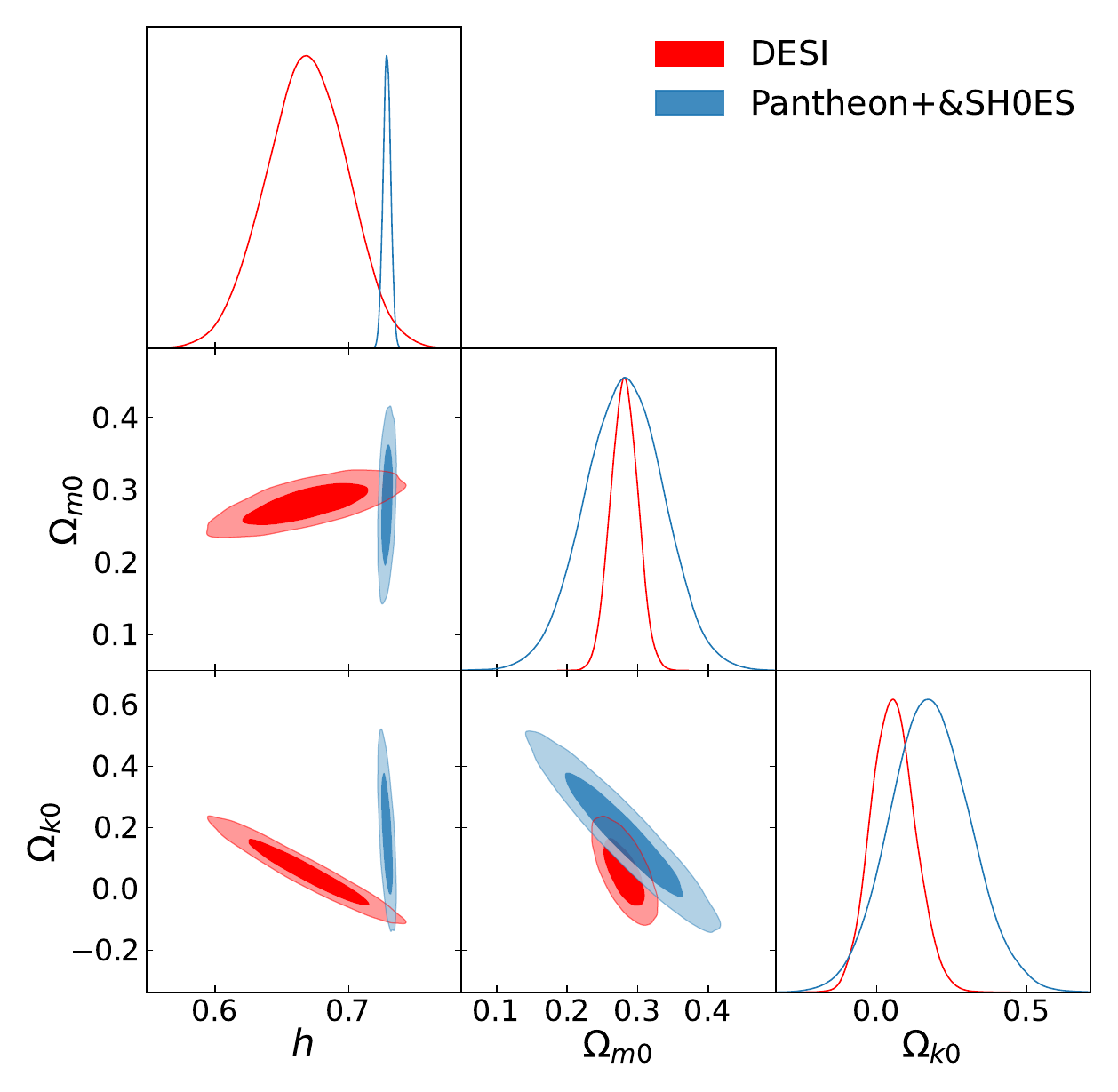}
       \includegraphics[width=\linewidth, height=0.5\linewidth]{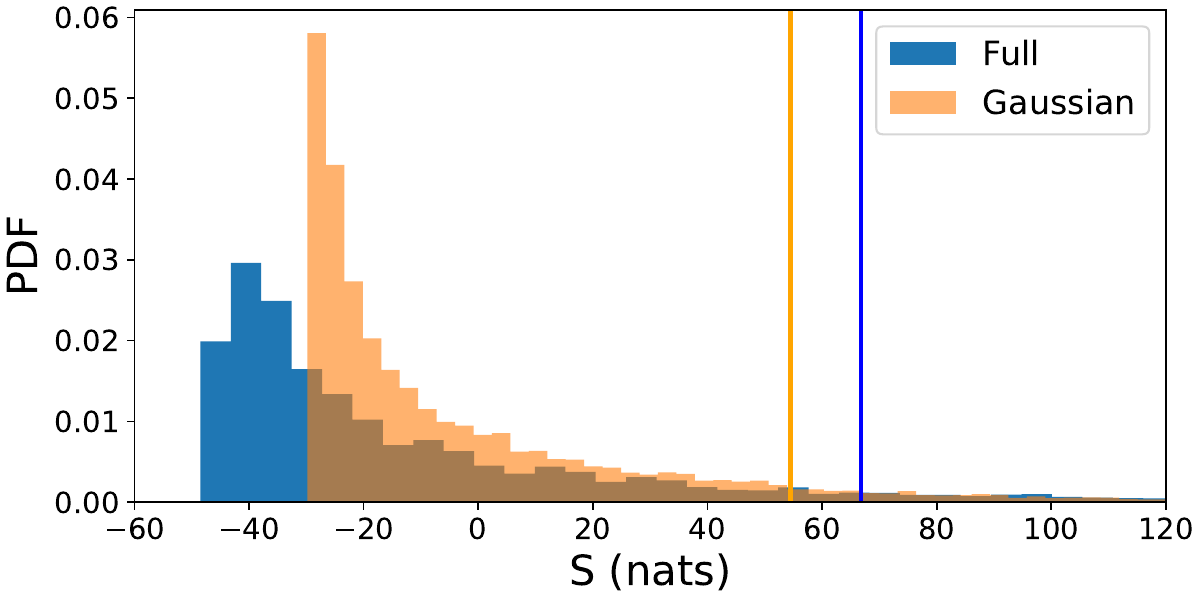}
    \end{minipage}
    \qquad
    \vspace{0.01em}
    \begin{minipage}{0.9\columnwidth}
        \includegraphics[width=\linewidth]{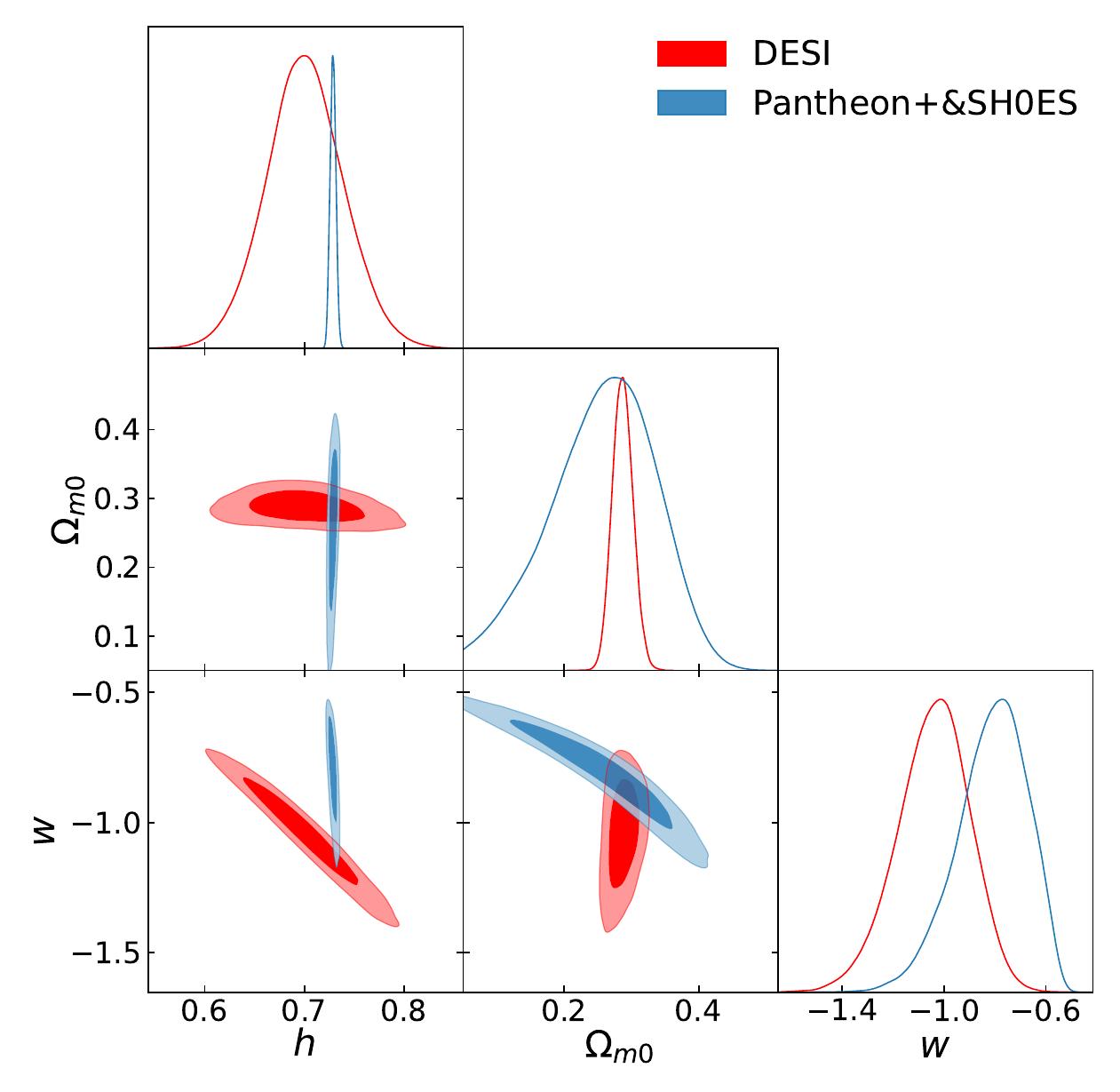}
        \includegraphics[width=\linewidth, height=0.5\linewidth]{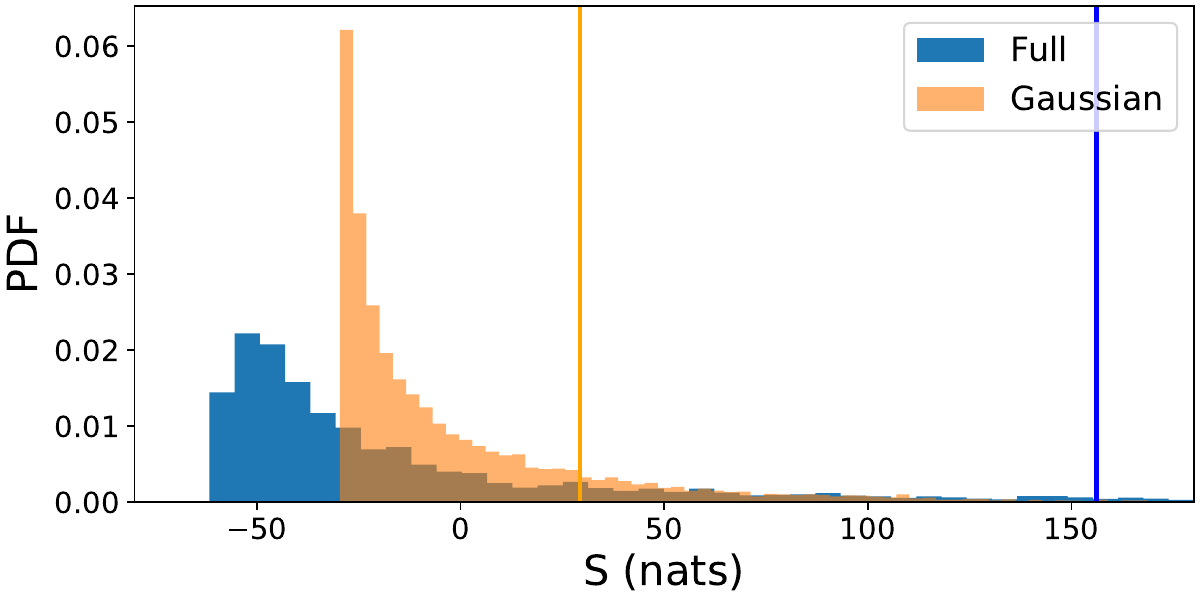}
        \centering
    \end{minipage}
    \caption{Same as Figures~\ref{fig:DESI_BOSS_Pantheon} and~\ref{fig:S-histograms} combined, but for the o$\Lambda$CDM (left panel) and the $w$CDM (right) models, and for the case where $\mathcal{D}_1$ = DESI and $\mathcal{D}_2$ = Pantheon+$\,$\&$\,$SH0ES. }
    \label{fig:SNIa_DESI_wCDM_oLCDM}
    \bigskip
\end{figure*}

\smallskip
\subsubsection{$S$ for the {\rm o$\Lambda$CDM} model}
A notably common extension of the $\Lambda$CDM model is the o$\Lambda$CDM, which posses a non-vanishing curvature parameter $\Omega_{k0}$. The contour plots of the distributions in the top-left panel of Figure~\ref{fig:SNIa_DESI_wCDM_oLCDM} show some resemblance to Gaussian distributions, the values for relative entropy  in Table~\ref{tab:BAO_results} suggest a slight deviation from Gaussianity, but less pronounced than in previous examples such as the o$w$CDM case. The numerical result for the Surprise distribution shown in the bottom-left panel of Figure~\ref{fig:SNIa_DESI_wCDM_oLCDM} deviates less from the analytical solution of the Surprise, shown in orange.

Besides the differences between the analytical and the numerical solution, the $p$-value indicates a similar significance between the results and indicate no tensions between neither BOSS+ or DESI and Pantheon+$\,$\&$\,$SH0ES.

\smallskip
\subsubsection{$S$ for the {\rm $w$CDM} model}

In this model, both all posterior distributions show clear signs of non-Gaussian behaviour, as can be seen in the top-right panel of Figure~\ref{fig:SNIa_DESI_wCDM_oLCDM}.

Their respective values of relative entropy, as can be seen in Table~\ref{tab:BAO_results}, are considerably different from their Gaussian counterparts.  The BOSS+ datasets presents no tension and is in agreement with the Pantheon+$\,$\&$\,$SH0ES dataset, both when considering the numerical and analytical Surprise statistics. The DESI dataset however presents a slight disagreement, with a $p$-value of 6\%, which is a tension of approximately $1.9\sigma$. This slight tension is only present when considering the full solution for the Surprise, and it is dissipated by the analytical treatment of the Surprise.

\smallskip
\subsubsection{$S$ for the flat-$\Lambda${\rm CDM} model}

The flat-$\Lambda$CDM ($\Lambda$CDM) model posses no curvature and a cosmological constant, whose equation of stare parameter is $w = -1$. In the F$\Lambda$CDM model, the posterior distributions are very well approximated by Gaussians, as can be seen by the results in Table~\ref{tab:BAO_results}. The numerical distribution for the Surprise  is the same as the analytical one, and the Gaussian Linear Model is a good approximation in this case. In this case, the Surprise points towards a slight tension of around 1.9$\sigma$ between $\mathcal{D}_2 = $ Pantheon+$\,$\&$\,$SH0ES and $\mathcal{D}_1 = $ BOSS+$\,$\&$\,$BBN. If we choose $\mathcal{D}_1 = $ DESI+BBN, however, there is a significant tension with a $p$-value of 0.08\%, this $p$-value corresponds to a 3.4$\sigma$ tension between datasets.

\smallskip
\subsubsection{Removing BBN and SH0ES data}

The tension between DESI and Panthon+ comes in part from the SH0ES and BBN data. To illustrate this, we make one further test. When computing the distribution of the Surprise, one can marginalize the posteriors over the parameter $h$ before computing the Kullback-Leibler divergence. This isolates the posterior from the effects of the SH0ES and BBN priors and provides a distribution of the Surprise without the parameter $h$. The results for the Surprise marginalizing over $h$ are presented in Table~\ref{tab:margh_full}. In this particular parameter space $\{\Omega_m, \Omega_k, w\}$, Pantheon+ and DESI are in better agreement for all four models here considered. In particular, for $\Lambda$CDM, the tension is alleviated from $3.4$ to $2.6\sigma$.

\begin{table}[t]
    \centering
    \begin{tabular}{ccccc}
        \hline\hline
        \multicolumn{5}{c}{\textbf{DESI vs. Pantheon+}} \\
        \hline
        & o$w$CDM & o$\Lambda$CDM & $w$CDM & $\Lambda$CDM \\
        \hline
        $D_{\rm KL}$ (nats)                 & 51.9  & 4.5   & 17.2  & 9.4  \\
        $\langle D_{\rm KL} \rangle$ (nats) & 46.9  & 8.8   & 20.7  & 1.4   \\
        $S$ (nats)                          & 5.4   & -4.3  & -3.5  & 7.9  \\
        $p$-value                           & 26\%  & 72\%  & 40\%  & 0.9\% \\
        Significance ($\sigma$)             & 1.1  & 0.4  & 0.8  & \high{2.6}  \\
        \hline\hline
    \end{tabular}
    \caption{Same as Table~\ref{tab:BAO_results} for DESI and Pantheon+ but without SH0ES.
    }
    \label{tab:margh_full}
    \smallskip
\end{table}

%%%%%%%%%%%%%%%%%%%%%%%%%%%%%%%%%%%%%%%%%%%%%%%%%%%%%%%%%%%%%%%%%%%%%%%%%%%%%%%%%%%%%%%%%%%%%%%%%%%%%%%%%%%%%%%%%

%%%%%%%%%%%%%%%%%%%%%%%%%%%%%%%%%%%%
\bigskip
\section{Conclusions}
\label{sec:conclusions}
%%%%%%%%%%%%%%%%%%%%%%%%%%%%%%%%%%%%

The Surprise is a powerful tool to quanitfy discrepancies in multi-dimensional parameter spaces. Analysing the Surprise statistic beyond its Gaussian approximation provides a more robust analysis of the tensions between various cosmological probes. We find significant deviations from the Gaussian approximation in some cases, particularly when considering models beyond the standard $\Lambda$CDM framework and using supernova data. We demonstrated with a mock data example how the non-Gaussian Surprise can successfully identify non-overlapping shapes in high dimensional space and point for discrepancies in data, which would otherwise be hidden using a Gaussian approximation with the full parameter covariance obtained with an MCMC chain.

We revealed a relevant $3.4\sigma$ tension between the Pantheon+$\,$\&$\,$SH0ES supernova data and  DESI BAO + BBN data if one assumes $\Lambda$CDM. This discrepancy, however, drops to $1.5-1.9\sigma$ in o$\Lambda$CDM, $w$CDM and o$w$CDM, which have more degrees of freedom.  We find that these discrepancies are in part driven by the tension in the Hubble constant measurements: removing the SH0ES data changes the  $\Lambda$CDM tension from $3.4\sigma$ to $2.6\sigma$. This result is in agreement with the main DESI results, which have found a preference for models with evolving dark energy over $\Lambda$CDM~\citep{DESI:2024mwx}.

Comparing instead Pantheon+$\,$\&$\,$SH0ES with BOSS+ BAO + BBN data, the discrepancies in $\Lambda$CDM are only at the 1.9$\sigma$ level, and remain at similar levels for the other three models, with a maximum value of $2.0\sigma$ for o$w$CDM.

The findings above are robust, and independent of assumption on the Gaussianity of the Surprise. In particular, the borderline $2.0\sigma$ tension between supernova and BOSS data for o$w$CDM is completely hidden if one takes the Gaussian approximation for the surprise.  Our results are also similar if one adopts either narrower or wider priors on the four cosmological parameters. The Surprise significance does depend in principle on the choice of priors, but if a wide enough prior is chosen, the discrepancy levels converge.

Some concerns have been raised in the literature that the conventional BAO analysis may contain a level of model dependence and underestimated uncertainties~\citep{Anselmi:2018vjz}. It could be interesting to investigate whether alternative analyses result in better agreement between DESI BAO and supernova data.

For BAO data alone, the Gaussian approximation gives reasonable results for the Surprise. For SNIa data instead, the Gaussian approximation is only accurate in the $\Lambda$CDM model. In the more general models, for which the parameters are less constrained, the distribution of the Surprise  deviates significantly from its Gaussian solution, and the Gaussian Surprise fails to capture the full extent of the discordance when supernova data were included.

The full non-Gaussian surprise calculations with the \texttt{klsurprise} code took around 500 CPU hours per dataset pair, which is a very moderate computational cost. This means that for small parameter spaces, one should perform the full calculations. In the future, we plan to investigate how the code computational cost scales with a larger parameter space.

\bigskip
\section*{Acknowledgements}

We thank Luca Amendola for useful discussions. PRM and MQ would like to express our sincere gratitude to Heidelberg University for their hospitality and support provided during our stay. MQ is supported by the Brazilian research agencies Fundação Carlos Chagas Filho de Amparo à Pesquisa do Estado do Rio de Janeiro (FAPERJ) project E-26/201.237/2022, CNPq (Conselho Nacional de Desenvolvimento Científico e Tecnológico) and CAPES (Coordenação de Aperfeiçoamento de Pessoal de Nível Superior). PRM is supported by the Brazilian research agency CAPES (Coordenação de Aperfeiçoamento de Pessoal de Nível Superior). BS is supported by Vector-Stiftung. We acknowledge the use of the computational resources of the joint CHE / Milliways cluster, supported by a FAPERJ grant E-26/210.130/2023.

%%%%%%%%%%%%%%%%%%%%%%%%%%%%%%%%%%%%%%%%%%%%%%%%%%
\bigskip
\section*{Data and Software Availability}

The data underlying this article will be shared on reasonable request to the corresponding author. The main \texttt{klsurprise} code is available at \url{https://github.com/pribamello/klsurprise}. We also provide, in the same link, interactive 3D plots of Figures~\ref{fig:SNIa_mocks_full} and \ref{fig:3D_contours} in Wolfram CDF format.

%%%%%%%%%%%%%%%%%%%% REFERENCES %%%%%%%%%%%%%%%%%%

\bigskip

\bibliographystyle{mn2e_eprint_new}
\bibliography{references}

%\normalsize

%%%%%%%%%%%%%%%%%%%%%%%%%%%%%%%%%%%%%%%%%%%%%%%%%%
%%%%%%%%%%%%%%%%% APPENDICES %%%%%%%%%%%%%%%%%%%%%

%\bigskip

\appendix

\medskip
\section{The choice of priors} \label{app:priors}

\begin{table*}[t!]
    \centering
    \begin{tabular}{ccccccc}
    \hline\hline
    & \multicolumn{3}{c}{{o$w$CDM}} & \multicolumn{3}{c}{{$\Lambda$CDM}} \\
    \hline
    & Default prior & {Wide prior} & {Narrow prior} &  Default prior & {Wide prior} & {Narrow prior} \\
    \hline
    \multicolumn{7}{c}{\textbf{BOSS+ vs. Pantheon+$\,$\&$\,$SH0ES}} \\
    \hline
    $D_{\rm KL}$ (nats)                 & 246    & 257    & 262    & 7.1   & 7.1    & 7.1 \\
    $\langle D_{\rm KL} \rangle$ (nats) & 113    & 120    & 148    & 3.9   & 3.8   & 3.2 \\
    S (nats)                            & 133    & 137    & 114    & 3.2   & 3.3   & 3.9 \\
    $p$-value                           & 4.5\% & 5.3\% & 17\%  & 6.1\% & 5.9\% & 4.1\% \\
    {significance ($\sigma$)}           & {2.0}   & 1.9   & 1.4   & 1.9   & 1.9   & 2.0 \\
    \hline
    \multicolumn{7}{c}{\textbf{DESI vs. Pantheon+$\,$\&$\,$SH0ES}}\\
    \hline
    $D_{\rm KL}$ (nats)                 & 334    & 351    & 303    & 18.5   & 18.5   & 18.5 \\
    $\langle D_{\rm KL} \rangle$ (nats) & 198    & 236    & 104    & 3.3   & 3.4   & 2.9 \\
    S (nats)                            & 136    & 115    & 199    & 15.1   & 15.0   & 15.6 \\
    $p$-value                           & 12\% & 18\% & 6.9\% & 0.10\% & 0.02\% & 0.08\% \\
    {significance ($\sigma$)}           & 1.5   & 1.3   & 1.8   & \high{3.3} & \high{3.7} & \high{3.4} \\
    \hline  \hline
    \end{tabular}
    \caption{Surprise  values for o$w$CDM and $\Lambda$CDM considering also a wider and a narrower set of priors. Values in red mark significances above $2\sigma$.}
    \label{tab:priors_results}
    \smallskip
\end{table*}

\begin{table}[t]
    \centering
    \begin{tabular}{cccc}
    \hline\hline
    & \multicolumn{2}{c}{Priors} \\ \hline
    Parameter  & Default prior       & Wide prior       & Narrow prior \\
    \hline
    $h$        & [0.3, 1.0]  & [0.3, 1.2]  & [0.2, 1]  \\
    $\Omega_m$ & [0.05, 1.0] & [0.05, 1.2] & [0.01, 0.99] \\
    $\Omega_k$ & [-1, 1]     & [-2, 2]     & [-0.3, 0.3] \\
    $w$        & [-3, -0.4]  & [-5, -0.1]  & [-3, 1] \\
    \hline  \hline
    \end{tabular}
    \caption{Priors for cosmological parameters used in the analysis. The Default priors are the ones used in the main text.}
    \label{tab:priors}
    \smallskip
\end{table}

One important choice when computing the Surprise statistics is the choice of prior. As seen in Figure~\ref{fig:3D_contours}, the marginalization of the parameters can hide the posterior mass in its projections. However, such an effect is not present when analyzing contours in more dimensions. This means that depending on the choice of priors, the posterior mass can be left unaccounted for and contribute significantly to the surprise. To study such effects, we chose the o$w$CDM and $\Lambda$CDM models and recomputed the Surprise statistics with a different prior range. Note from the results in Table~\ref{tab:DESI_SDSS_Pantheon_DESI_results} that when going from a less restrictive o$w$CDM to a more restrictive model $\Lambda$CDM, counterintuitively the tensions increase. To see if this can be a prior induced effect, we chose these two specific models to recompute the Surprise statistics.
The results for all three priors are shown in Table~\ref{tab:priors_results}. The details of all three priors are shown in Table~\ref{tab:priors}.

Note that the pattern of increasing tension when changing to a more constraining cosmological model remains in different choices of priors. This indicates that such effects are not caused by the choice of priors. In fact, different priors do not significantly change the results in $\Lambda$CDM, which is expected, as the confidence contours in this model with less degrees of freedom are fully contained by the likelihood itself. Different priors also cause little change in the results for the o$w$CDM case.

\medskip
\section{Convergence of KLD} \label{app:kld_convergence}
To assess the convergence of our KLD estimate we performed multiple calculations of KLD for the same dataset varying the number of samples used to compute the integral in Eq.~\eqref{eq:kld_numerical}. For each number of samples, we evaluated KLD one thousand times and compared the distributions we obtained. Finally, we opted for a middle ground that preserves precision and does not increase the computational time significantly.

To be more precise, let's consider the case of the o$w$CDM model with $\mathcal{D}_2 = $ Pantheon+$\,$\&$\,$SH0ES and $\mathcal{D}_1$. We start by choosing a number of samples that will be used to compute KLD with Eq.~\eqref{eq:kld_numerical}. We then perform the sampling of $p(\theta\mid\mathcal{D}_i)$ using Nested Sampling. We force the number of effective samples, compute KLD by means of Eq.~\eqref{eq:kld_numerical} and, finally, repeat the process one thousand times for each multiple values of effective samples. This ensures that we have distributions of values of KLD, which can be used as an estimate of the code accuracy.

\begin{figure}[t]
    \centering
    \includegraphics[width=.93\linewidth]{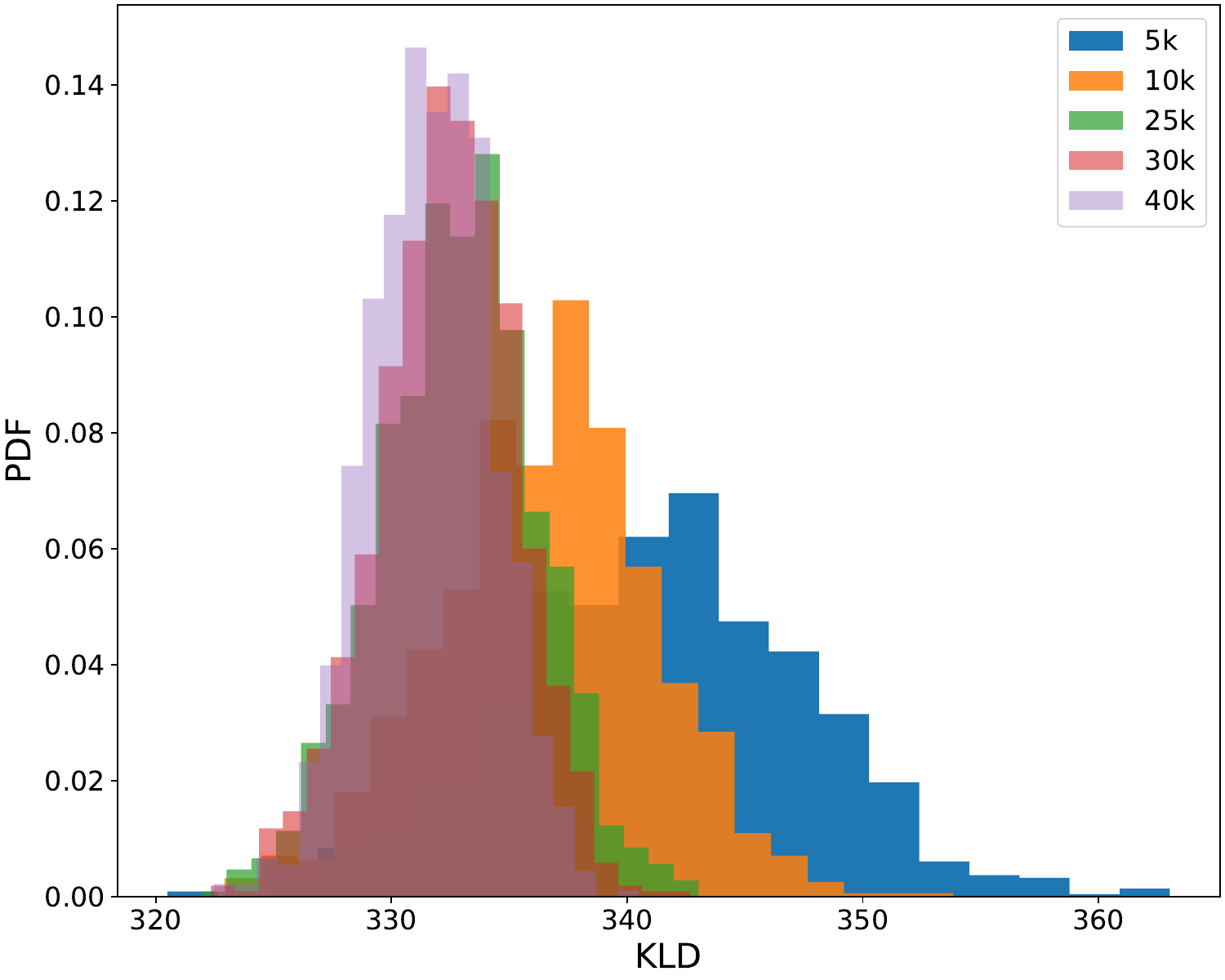}
    \caption{The distribution of KLD accordingly to the number of samples in Eq.~\eqref{eq:kld_numerical}. The number of effective samples used in this paper was of 25000.}
    \label{fig:kld-convergence}
    \medskip
\end{figure}

Figure \ref{fig:kld-convergence} shows how the distribution of KLD changes according to the number of samples used to compute Eq.~\eqref{eq:kld_numerical}. As can be seen, results converge around 25000 samples. For this reason, this was our choice for number of posterior samples. We infer that the our KLD values have a precision of 1\%.

\medskip
\section{The klsurprise code} \label{app:code}

\begin{figure}[t]
    \centering
    \includegraphics[width=.95\linewidth]{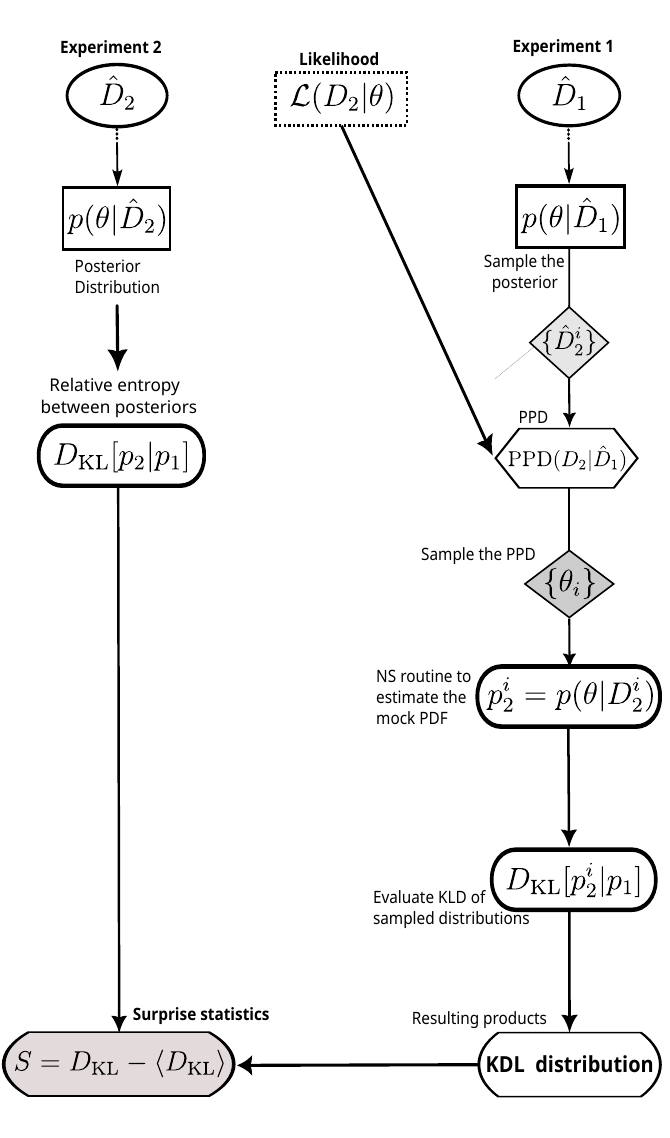}
    \caption{The main steps involved in evaluating the Surprise statistic distribution numerically.}
    \label{fig:S-flowchart}
    \medskip
\end{figure}

We built a python code to evaluate the Surprise  distribution. A flow chart description of this code is shown in Figure~\ref{fig:S-flowchart}. The code makes use of four main quantities, a python callable function of the posterior distribution $p(\Theta|\mathcal{D}_1)$ that takes as input a $N_p$-dimensional parameter vector $\Theta$, a python function which maps parameter space to a data model of likelihood 2, $\mathcal{D}_2(\theta)$, the covariance matrix $\Sigma_2$ and a domain, which is used internally as a flat uninformative prior on the parameters. The code has two main routines, the \texttt{create\_ppd\_chain} and the \texttt{compute\_kld\_distribution} routine.

The \texttt{create\_ppd\_chain} routine is the first step into computing the Surprise distribution. It rewrites the PPD in equations \eqref{eq:PPD} as \eqref{eq:PPD_numerical} and sample the distribution. The function takes as input a $\{N_s, d\}$ matrix, which we dub the $\Theta_{\rm M}$-matrix, containing the parameter samples $\{\Theta_i\}$ from the posterior distribution $p(\Theta|\mathcal{D}_1)$, where $N_s$ is the number of samples of the integral, as per Eq.~\eqref{eq:PPD_numerical}, and $d$ is the dimensionality of the parameter space. Another important parameter for the \texttt{create\_ppd\_chain} routine is the covariance matrix $\Sigma_2$ from the likelihood function $\mathcal{L}(\mathcal{D}_2|\Theta)$.

The $\Theta_{\rm M}$ parameter matrix can then be converted into a mock data matrix $\mathcal{D}_2^{\rm mock}(\Theta)$ of dimensionality $\{N_s, N\}$, where $N$ is the dimensionality of data space. This matrix is then used to build $N_s$ likelihood functions $\{\mathcal{L}(\mathcal{D}_2^{\rm mock}|\Theta)\}$ which are then used to sample the PPD distribution in Eq.~\eqref{eq:PPD_numerical}. The sampling process could be done in two ways. If the dimensionality of data-space $N$ is small enough, one can write Eq.~\eqref{eq:PPD_numerical} as a callable function of $\mathcal{D}_2$ and use any preferred sampling technique (such as MCMC or nested sampling). Alternatively, if the dimensionality of the data-space is prohibitively large, as in the case of a SNIa dataset, then one can randomly choose a subset of the mock likelihood functions $\{\mathcal{L}(\mathcal{D}_2^{\rm mock}|\Theta)\}$. In this case, we assume them to be Gaussian, which for the case of a SNIa dataset is a reasonable assumption, sample the mock likelihoods and take all samples as the description of the PPD. Some tricks can be used to make the sampling process faster, such as Cholesky decomposition of the covariance matrix $\Sigma_2$ for instance. The samples of the PPD will induce the distribution of KLD.

For the second sampling method, convergence is assured if the sampling of each individual likelihood in Eq.~\eqref{eq:PPD_numerical} has converged. As the PPD is thus written as a sum of likelihoods, by assigning equal weights to each, and ensuring that their sampling has converged, we can also guarantee that the sampling of the PPD has converged. The number of samples from the PPD can then be taken according to the number of desired samples of $S$ in the final distribution.

The routine \texttt{compute\_kld\_distribution} takes a matrix of dimensionality $\{N_{\rm ppd}, N\}$, where $N_{\rm ppd}$ is the number of PPD samples and each line of size $N$ is a PPD sample. $N_{\rm ppd}$ has to be large enough to ensure a good resolution of the surprise $p$-value, and, in case the PPD sampling is performed with an MCMC or nested sampling technique, also large enough to ensure convergence of the sample.
\texttt{compute\_kld\_distribution} returns a vector of $N_{\rm ppd}$ values for the KLD between the posterior constructed with each PPD sample of $\mathcal{D}_2^{\rm mock}(\Theta)$ and the posterior distribution $p(\Theta|D_1)$. This function iteratively runs a Nested Sampling routine to estimate the posterior distributions from the $i{\rm th}$ entry of $\mathcal{D}_2^{\rm mock}(\Theta)$, $p(\Theta|D_{2,i}^{\rm mock})$, and then compute the value of KLD using Eq.~\eqref{eq:kld_numerical}. We use Nested Sampling to reconstruct the posteriors needed for computing this integral~\citep{Skilling:2004pqw,Skilling:2006gxv}, specifically the python Dynesty implementation \citep{2020MNRAS.493.3132S}. We  make use of both the static and Dynamic Nested Sampling implementation \citep{Higson_2018} so we can properly sample the parameter space and get a good estimate of the evidence $Z$. The evidence estimation is a fundamental step into computing the value of KLD in Eq.~\eqref{eq:kld_numerical}, as both the posterior used must be normalized.

One could for instance try to avoid Nested Sampling altogether, use a preferred Monte Carlo technique to sample the posterior distribution and then estimate their normalized densities by means of a Kernel Density Estimator approach, but that will likely ultimately fail, as the reconstruction of the posterior distributions used in Eq.~\eqref{eq:kld_numerical} must be also very accurate even in the regions with low posterior values, as these can be important when evaluating the numerical KLD of two disagreeing distributions.

The \texttt{compute\_kld\_distribution} routine will result in a vector $\{D_{\rm KL}^i \}$ with  $N_{\rm ppd}$ values for $D_{\rm KL}$ constructed with samples from the PPD. These are then used to reconstruct the distribution of KLD and consequently, the distribution of the Surprise. This routine is the main bottleneck of the code, as constructing the posterior distribution for each PPD sample is a time-consuming process and scales accordingly to the dimensionality of parameter space and the efficiency of the Nested Sampling algorithm used. One can then use the routine  \texttt{run\_nested\_sampling} and \texttt{KLD\_numerical} to find the value of KLD between the two posterior distributions $p(\Theta|\mathcal{D}_1)$ and $p(\Theta|\mathcal{D}_2)$. Finally, the value of $\langle D_{\rm KL} \rangle$ can be found by averaging the resulting vector of \texttt{compute\_kld\_distribution} $\{D_{\rm KL}^i \}$. The Surprise is found by using Eq.~\eqref{eq:surprise} and its $p$-value is then evaluated from both $S$ and its distribution. As the process of evaluating the distribution of $S$ mixes a Bayesian and a Frequentist approach, the resulting $p$-value will have its error limited by the number of resulting samples in the Surprise distribution.

The most time-consuming part of the code, to wit computing the KLD distribution, can be easily parallelized. We use \texttt{joblib} to distribute the computation of the different Nested Sampling runs needed to estimate $p(\Theta|\mathcal{D}_2^j)$ for each PPD sample. With the reconstructed PPD mock posteriors, the integral in Eq.~\eqref{eq:kld_numerical}  can be easily vectorized by using Jax library \citep{jax2018github}. Sampling the PPD itself is not time-consuming, usually taking less than 1 minute for the quantities of samples we need. The plots in this paper were created with 5000 samples of KLD. For the case with $\mathcal{D}_1 = $ BOSS+$\,$\&$\,$BBN and $\mathcal{D}_2 = $ Pantheon+$\,$\&$\,$SH0ES, the total computation time was 540 CPU hours (wall time around 11h in our 50-core workstation). If we consider a different scenario with $\mathcal{D}_1 = $ BOSS+ and $\mathcal{D}_2 = $ DESI, the computational times shrink significantly, with the whole code taking about 71 CPU hours (wall time around 1h30). Ultimately, the run times will vary significantly, depending also on the likelihood used and its implementation. It will increase linearly with the number of points one needs for the $D_{\rm KL}$ distribution.

\bigskip
\pagebreak

\label{lastpage}

\end{document}